\documentclass[a4paper,fleqn,usenatbib]{mnras}

\usepackage{newtxtext,newtxmath}

\usepackage[T1]{fontenc}
\usepackage{ae,aecompl}

\usepackage{graphicx}	
\usepackage{amsmath}	
\usepackage{amssymb}	
\usepackage[flushleft]{threeparttable}
\usepackage{comment}
\usepackage{multirow}

\newcommand{\flux}{\,erg\,cm$^{-2}$\,s$^{-1}$}

\title[XUV environments of exoplanets]{The XUV environments of exoplanets from Jupiter-size to super-Earth}

\author[G. W. King et al.]{George W. King,$^{1,2}$\thanks{Email: g.king.5@warwick.ac.uk}
Peter J. Wheatley,$^{1,2}$\thanks{Email: p.j.wheatley@warwick.ac.uk}
Michael Salz,$^{3}$
Vincent Bourrier,$^{4}$
\newauthor
Stefan Czesla,$^{3}$
David Ehrenreich,$^{4}$
James Kirk,$^{1}$
Alain Lecavelier des Etangs,$^{5}$
\newauthor
Tom Louden,$^{1,2}$
J\"{u}rgen Schmitt,$^{3}$
and P. Christian Schneider$^{3}$
\\
$^{1}$Department of Physics, University of Warwick, Gibbet Hill Road, Coventry, CV4 7AL, UK\\
$^{2}$Centre for Exoplanets and Habitability, University of Warwick, Gibbet Hill Road, Coventry, CV4 7AL, UK\\
$^{3}$Hamburger Sternwarte, Universit\"{a}t Hamburg, Gojenbergsweg 112, 21029 Hamburg, Germany\\
$^{4}$Observatoire de l'Universit\'{e} de Gen\`{e}ve, 51 chemin des Maillettes, 1290 Sauverny, Switzerland\\
$^{5}$Institut d'Astrophysique de Paris, UMR7095 CNRS, Universit\'{e} Pierre \& Marie Curie, 98bis boulevard Arago, 75014 Paris, France
}

\date{Accepted XXX. Received YYY; in original form ZZZ}

\pubyear{2017}

\begin{document}
\label{firstpage}
\pagerange{\pageref{firstpage}--\pageref{lastpage}}
\maketitle

\begin{abstract}
Planets that reside close-in to their host star are subject to intense high-energy irradiation. Extreme-ultraviolet (EUV) and X-ray radiation (together, XUV) is thought to drive mass loss from planets with volatile envelopes. We present \textit{XMM-Newton} observations of six nearby stars hosting transiting planets in tight orbits (with orbital period, $P_\text{orb} < 10$\,d), wherein we characterise the XUV emission from the stars and subsequent irradiation levels at the planets. In order to reconstruct the unobservable EUV emission, we derive a new set of relations from Solar \textit{TIMED/SEE} data that are applicable to the standard bands of the current generation of X-ray instruments. From our sample, WASP-80b and HD\,149026b experience the highest irradiation level, but HAT-P-11b is probably the best candidate for Ly\,$\alpha$ evaporation investigations because of the system's proximity to the Solar System. The four smallest planets have likely lost a greater percentage of their mass over their lives than their larger counterparts. We also detect the transit of WASP-80b in the near ultraviolet with the Optical Monitor on \textit{XMM-Newton}.\\
\end{abstract}




\section{Introduction}

A substantial number of the exoplanets discovered to date have orbital periods of less than 10 days, lying much closer to their host star than Mercury does the Sun. Such close-in planets are subject to strong irradiation by their parent star. High-energy photons at extreme-ultraviolet and X-ray wavelengths are thought to drive hydrodynamic winds from planetary atmospheres. It has been suggested that super-Earth and Neptune-sized planets may be susceptible to losing a significant portion of their mass through this process~\citep[e.g.][]{Owen2012,Lopez2013}. In the most extreme cases, complete atmospheric evaporation and evolution to a largely rocky planet may be possible. This is an area of particular interest given past studies that point to a dearth of hot Neptunes at the very shortest periods, where high-energy irradiation is at its greatest~\citep[e.g.][]{LDE2007,Davis2009,Ehrenreich2011,Szabo2011,Beauge2013,Helled2016}. This shortage cannot be explained by selection effects. In contrast, hot Jupiters should lose only a few percent of their envelope on timescales of the order of Gyr~\citep[e.g.][]{MurrayClay2009,Owen2012}.

Detections of substantial atmospheric expansion and mass loss have been inferred in a few cases as significant increases in the observed transit depth in ultraviolet lines, particularly around the Ly\,$\alpha$ line which probes neutral hydrogen. The first such inference was made by \citet{VidalMadjar2003}, who detected in-transit absorption in the Ly\,$\alpha$ line ten times that expected from the optical transit of HD\,209458b. This depth pointed to absorption from a region larger than the planet's Roche lobe, leading to the conclusion of an evaporating atmosphere. The light curve also showed an early ingress and late egress. Ly\,$\alpha$ observations of HD\,189733b also revealed an expanded atmosphere of escaping material~\citep{LDE2010,LDE2012,Bourrier2013}. The measured Ly\,$\alpha$ transit depth was measured at twice the optical depth in 2007, and six times the optical depth in 2011. The temporal variations in the transit absorption depth of the Ly\,$\alpha$ line may be related to an X-ray flare detected in contemporaneous \textit{Swift} observations in 2011. Similar signatures are beginning to be detected for sub Jovian sized planets too. \citet{Ehrenreich2015} discovered an exceptionally deep Ly\,$\alpha$ transit for GJ\,436b, implying that over half of the stellar disc was being eclipsed. This is compared to the 0.69 per cent transit seen in the optical. The Ly\,$\alpha$ transits show early ingress, and late egress up to 20 hours after the planet's transit~\citep{Lavie2017}. Recent investigations by~\citet{Bourrier2017} for the super-Earth HD\,97658b suggested that it does not have an extended, evaporating hydrogen atmosphere, in contrast to predictions of observable escaping hydrogen due to the dissociation of steam. Possible explanations include a relatively low XUV irradiation, or a high-weight atmosphere. A similar non-detection for 55\,Cnc\,e had previously been found by~\citet{Ehrenreich2012}. In that case, complete loss of the atmosphere down to a rocky core has not been ruled out. Most recently, a study of Kepler-444 showed strong variations in the Ly\,$\alpha$ line that could arise from hydrodynamic escape of the atmosphere~\citep{Bourrier2017K}, while Ly\,$\alpha$ emission has been detected for TRAPPIST-1~\citep{Bourrier2017T} and the search for evaporation signatures is ongoing.

Evidence for hydrodynamic escape also extends to lines of other elements. Deep carbon, oxygen and silicon features have been detected for HD\,209458b, and interpreted as originating in the extended atmosphere surrounding the planet~\citep{VidalMadjar2004,Linsky2010}, the species having been entrained in the hydrodynamic flow. The Si III detection in this case has since been found to be likely due to stellar variability~\citep{Ballester2015}. A similar O\,\textsc{I} detection has been made for HD\,189733b~\citep{BenJaffel2013}. 

EUV photons are thought to be an important driving force behind atmospheric evaporation~\citep[e.g.][]{Owen2012}, however they are readily absorbed by the interstellar medium, making direct observations possible only for the closest and brightest stars. Additionally, since the end of the EUVE mission~\citep{Bowyer1991} in 2001, no instrument covers this spectral range, so the EUV emission of host stars can only be derived from estimates. \citet{LDE2007} and~\citet{Ehrenreich2011} applied a method of reconstruction from a star's rotational velocity, using a relationship from the work of~\citet{Wood1994}, and several studies have looked at linking EUV output to observable wavelengths. \citet{SanzForcada2011} derived an expression relating the EUV and X-ray luminosities, based on synthetic coronal models for a sample of main sequence stars. \citet{Chadney2015} analysed Solar data from the \textit{TIMED/SEE} mission~\citep{Woods2005}, determining an empirical power law relation between the ratio of EUV to X-ray flux, and the X-ray flux at the stellar surface. Their fig.~2 suggests that this relation appears consistent with measurements of a small sample of nearby stars. \citet{Linsky2014} approached the problem from the other direction, reconstructing the EUV emission from Ly\,$\alpha$. They combined Ly\,$\alpha$ observations with EUVE measurements in the range 100 to 400\,\AA, and solar models from~\citet{Fontenla2014} across 400 to 912\,\AA. This method has since been employed by the MUSCLES Treasury Survey, who have catalogued SEDs for 11 late type planet hosts from X-ray through to mid-infrared~\citep{Youngblood2016,Loyd2016}. An alternative approach is to perform a Differential Emission Measure recovery, as~\citet{Louden2017} did for HD\,209458. This study incorporated information from both the \textit{Hubble Space Telescope} and \textit{XMM-Newton} on the UV line and X-ray fluxes, respectively.

Following on from~\citet{Salz2015}'s investigations into hot Jupiters, we probe the high-energy environments of planets ranging from Jupiter-size down to super-Earth. All six planets in our sample orbit their parent star with a period of <10\,d. Our sample is introduced in Section~\ref{sec:Samp}. The observations are described in Section~\ref{sec:Obs}. Results of the X-ray analysis, as well as reconstruction of the full XUV flux are presented in Section~\ref{sec:Xresults}. The optical monitor results are detailed in Section~\ref{sec:OMres}. The implications of our analysis are discussed in Section~\ref{sec:Discuss}. The work is summarised in Section~\ref{sec:conclusions}.

\section{Sample}
\label{sec:Samp}

\begin{figure}
 \includegraphics[width=\columnwidth]{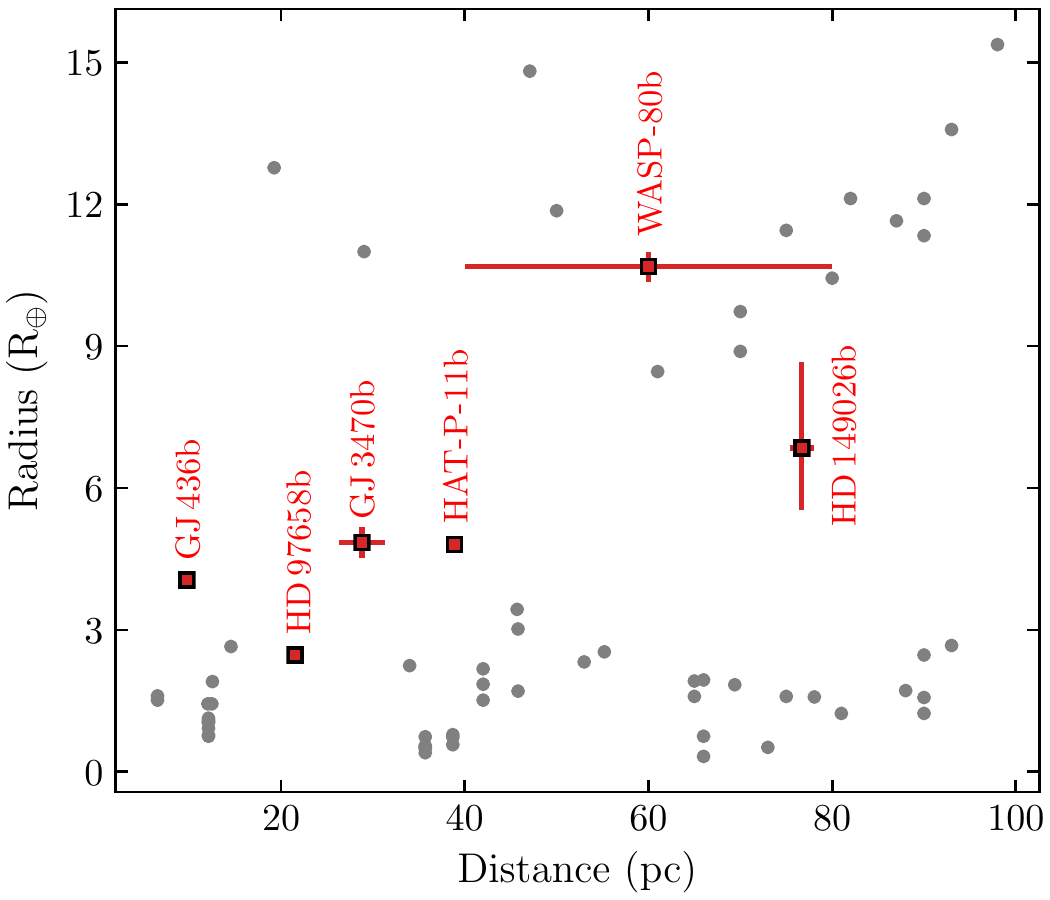}
 \caption{Distances and radii of all known transiting planets within 100\,pc of Earth. The six planets in our sample are shown as red squares. Other planets are shown as grey circles. Data taken from NASA Exoplanet Archive.}
 \label{fig:sample}
\end{figure}

\begin{table*}
	\centering
	\caption{System parameters for the six transiting exoplanet host stars we observed with \textit{XMM-Newton}.}
	\label{tab:SysParam}
	\begin{threeparttable}
	\begin{tabular}{lccccccc|ccccccc} 
		\hline
		System & Spectral & $V$ & $d$ & Age & $R_*$ & $T_\text{eff,\,*}$ & $P_{\text{rot}}$ & $R_\text{p}$ & $M_\text{p}$ & $\log\,g$ & $P_{\text{orb}}$ & $a$ & $e$ & $T_\text{eff,\,p}$ \\
		 & Type & (mag) & (pc) & (Gyr) & (R$_\odot$) & (K) & (d) & (R$_J$) & (M$_J$) & (cm/s$^2$) & (d) & (au) &  & (K) \\
		\hline
		GJ\,436 & M2.5V & 10.6 & 9.749 & 6 & 0.437 & 3585 & 44.09 & 0.361 & 0.0737 & 3.15 & 2.644 & 0.0287 & 0.153 & 740\\
		GJ\,3470 & M1.5V & 12.3 & 28.82 & 1--4 & 0.568 & 3600 & 20.7 & 0.432 & 0.0437 & 2.76 & 3.337 & 0.0369 & 0 & 620 \\
		HAT-P-11 & K4V & 9.5 & 37.88 & 5.2 & 0.752 & 4780 & 29.33 & 0.428 & 0.081 & 3.05 & 4.888 & 0.0513 & 0.2646 & 880 \\
		HD\,97658 & K1V & 7.7 & 21.53 & 9.7 & 0.741 & 5170 & 38.5 & 0.220 & 0.0238 & 3.17 & 9.489 & 0.080 & 0.078 & 760 \\
		HD\,149026 & G0IV & 8.1 & 76.7 & 1.2 & 1.290 & 6147 & 11.5 & 0.610 & 0.356 & 3.37 & 2.876 & 0.0429 & 0 & 1600 \\
		WASP-80 & K7-M0V & 11.9 & 60 & 0.1 & 0.571 & 4145 & 8.5 & 0.952 & 0.554 & 3.18 & 3.068 & 0.0346 & <0.07 & 800\\
		\hline
	\end{tabular}
    \begin{tablenotes}
\item References: GJ 436: All parameters from~\citet{Knutson2011} except $d$~\citep[\textit{Gaia}:][]{GaiaDR1}, age and $T_\text{eff,\,*}$~\citep{Torres2007}, $P_{\text{rot}}$~\citep{Bourrier2018}, and $M_\text{p}$~\citep{Southworth2010}. GJ 3470: All from~\citet{Awiphan2016}, except $d$ and $P_{\text{rot}}$~\citep{Biddle2014}. HAT-P-11: $R_*$, $M_\text{p}$, and $T_\text{eff,\,p}$ from~\citet{Bakos2010}, $R_\text{p}$, $P_\text{orb}$, $a$, and $e$ from~\citet{Huber2017,Huber2017C}, $d$ from \textit{Gaia}, age from~\citet{Bonfanti2016}, $P_{\text{rot}}$ from~\citet{Beky2014}. HD 97658: All from~\citet{VanGrootel2014}, except $d$ (\textit{Gaia}), age~\citep{Bonfanti2016}, $P_{\text{rot}}$~\citep{Henry2011}, $R_\text{p}$ and $P_{\text{orb}}$~\citep{Knutson2014}. HD 149026: All from~\citet{Southworth2010} ($P_{\text{rot}}$ from $v\sin i$), except $d$ (\textit{Gaia}), and $T_\text{eff,\,*}$~\citep{Sato2005}. WASP-80: All from~\citet{Triaud2013}, except $P_\text{orb}$~\citep{Mancini2014} and $P_{\text{rot}}$ from $v\sin i$~\citep{Triaud2015}.
    \end{tablenotes}
  \end{threeparttable}
\end{table*}

Our sample of six systems is made up of six of the closest known transiting planets to Earth, and is listed in Table~\ref{tab:SysParam}. Fig.~\ref{fig:sample} shows all known transiting planets within 100\,pc, and highlights the objects in this sample. Each of the planets occupies a scarcely populated area of this parameter space. Together with the results of past observations with \textit{XMM-Newton} and \textit{ROSAT} for some of the sample, their proximity means that all of the hosts were predicted to exhibit sufficient X-ray flux for characterisation of the planet's XUV irradiation.

Table~\ref{tab:SysParam} outlines the properties of each planetary system investigated. We note that the values for HD\,149026 from \citet{Southworth2010} differ substantially from those of~\citet{Carter2009}, and that this also affects our mass loss analysis in Section~\ref{ssec:MassLoss}.

GJ\,436b, GJ\,3470b, and HAT-P-11b are the three closest transiting Neptune-sized planets. Only two other transiting planets within 100\,pc, K2-25b and HD\,3167c, have a radius between 3 and 5\,R$_\oplus$, and the latter is likely to be comparatively far less irradiated than our sample. HD\,149026 is the only known exoplanet within 100\,pc with its radius between that of Neptune and Saturn. HD\,97658b is the second-closest, and orbits by far the brightest star ($V=7.7$\,mag) of any known planet of its size. Though its importance is less obvious from Fig.~\ref{fig:sample}, WASP-80 represents one of only a handful of transiting hot Jupiter's in orbit around a late K/early M-type star.

In addition to the favourable X-ray characterisation potential, four of the systems (GJ\,436, HAT-P-11, HD\,97658, and WASP-80) were also chosen in order to explore their near ultraviolet (NUV) transit properties with the Optical Monitor (OM) on \textit{XMM-Newton}.

\section{Observations}
\label{sec:Obs}

\begin{table*}
  \caption{Details of our \textit{XMM-Newton} observations.}
  \label{tab:Obs}
  \centering
  \begin{threeparttable}
  \begin{tabular}{l c c c c c c c c c c}
    \hline\
    Target & ObsID & PI & Start time & Exp. T & Start -- Stop  & Transit & PN & OM & Ref.\\
           &       &  & (TDB)           & (ks)     & phase & phase & filter & filter(s) & \\
    \hline
    GJ\,3470   & 0763460201 & Salz & 2015-04-15 03:13 & 15.0 & 0.838 -- 0.890 & 0.988 -- 1.012 & Medium & U/UVW1/UVM2 & 1 \\
    WASP-80   & 0764100801 & Wheatley & 2015-05-13 13:08 & 30.0 & 0.944 -- 1.065 & 0.986 -- 1.014 & Thin & UVW1 & 2 \\
    HAT-P-11  & 0764100701 & Wheatley & 2015-05-19 13:13 & 28.5 & 0.967 -- 1.035 & 0.990 -- 1.010 & Thin & UVW2 & 3 \\
    HD\,97658  & 0764100601 & Wheatley & 2015-06-04 04:35 & 30.9 & 0.980 -- 1.019 & 0.994 -- 1.006 & Medium & UVW2 & 4 \\
    HD\,149026 & 0763460301 & Salz & 2015-08-14 19:19 & 16.7 & 1.009 -- 1.077 & 0.977 -- 1.023 & Medium & UVM2/UVW2 & 5 \\
    GJ\,436    & 0764100501 & Wheatley & 2015-11-21 01:40 & 24.0 & 0.949 -- 1.063 & 0.992 -- 1.008 & Thin & UVW1 & 6 \\

    \hline
\end{tabular}
    \begin{tablenotes}
\item Start time and duration are given for EPIC-pn.
\item References for the ephemerides:
			(1) \citet{Biddle2014};
			(2) \citet{Triaud2013};
			(3) \citet{Huber2017C};
            (4) \citet{Knutson2014};
            (5) \citet{Carter2009};
            (6) \citet{Lanotte2014}.
    \end{tablenotes}
  \end{threeparttable}
\end{table*}

The six planet hosts were all observed by the European Photon Imaging Camera (EPIC) on \textit{XMM-Newton} in 2015. Table~\ref{tab:Obs} provides details of the observations in time, duration and orbital phase, as well as the adopted ephemerides. Observations were taken with the OM concurrently, cycling through different filters for GJ\,3470 and HD\,149026. For the other four objects, a single filter was used in fast mode in an attempt to detect transits in the ultraviolet.

The data were reduced using the Scientific Analysis System (\textsc{sas} 15.0.0) following the standard procedure\footnote{As outlined on the `SAS Threads' webpages: \url{http://www.cosmos.esa.int/web/xmm-newton/sas-threads}}. The EPIC-pn data of all systems except HD\,97658 show elevated high-energy background levels at some points in the observations. To minimise loss of exposure time, we raised the default count rate threshold for time filtering due to high-energy events (> 10\,keV) by a factor of two compared to the standard value. Background filtering does not affect the results, except for HAT-P-11. High background (exceeding this higher threshold) was observed at numerous epochs in the HAT-P-11 data, as often seen in \textit{XMM-Newton} due to Solar soft protons~\citep{Walsh2014}. Although the size of the uncertainties were not significantly changed by filtering, a 10 per cent increase in the best fit flux values were obtained with the filtered dataset. The results presented here use the filtered dataset in the spectral fitting process and subsequent analysis, however the light curve for HAT-P-11 presented in Section~\ref{ssec:XLC} uses the unfiltered dataset in order to avoid large gaps.

\subsection{Nearby sources}

HD\,149026 is not known to be a double star system \citep{Raghavan2006, Bergfors2013}, but STScI Digitized Sky Survey (DSS) images show a nearby star at 20\,arcsec distance north-east of HD\,149026. 
The source is also present in 2MASS images \citep{Skrutskie2006} and our OM data. In the multi-epoch DSS images HD\,149026 displays a proper motion of 3.651\,arcsec measured over a time period of 40 years \citep{Raghavan2006}. The nearby source is not co-moving, hence, we identified it as a background source.

DSS and 2MASS images contain a source 8\,arcsec away from HAT-P-11. This object was identified as KOI-1289 by the \textit{Kepler} mission, and later found to be a false positive due to a blended signal from HAT-P-11\footnote{Flagged as a false positive on the MAST Kepler archive: \url{https://archive.stsci.edu/kepler/}. Inspection of the light curves reveal a transit signal with the same period as HAT-P-11.}. KOI-1289 is 4.6\,mag fainter than HAT-P-11 in the B band, and 6.3\,mag fainter in the R band~\citep{Cutri2003,Hog2000,Monet2003}. This is consistent with our findings: KOI-1289 is barely detected in OM. Comparison of the OM positions of both objects to their respective J2000 positions~\citep{Cutri2003,vanLeeuwen2007} reveals proper motion in different directions at different rates, and are thus not co-moving.

WASP-80 also has a much fainter star located nearby (9\,arcsec), as discussed in section 3.1 of \citet{Salz2015}. They identify it as a background 2MASS source, 4\,mag dimmer than WASP-80.

In all four cases, the detected X-rays are centred on the exoplanet host star, and there is no evidence for X-rays from the nearby object. They were therefore neglected in the following analysis.

\section{X-ray analysis and results}
\label{sec:Xresults}

An X-ray source was detected within 1.5\,arcsec of the expected position of each target star. 15\,arcsec extraction regions were used for all sources, with multiple circular regions on the same CCD chip used for background extraction, located as close to the source as possible beyond 30\,arcsec.

\subsection{X-ray light curves}
\label{ssec:XLC}

\begin{figure*}
\centering
 \includegraphics[width=\textwidth]{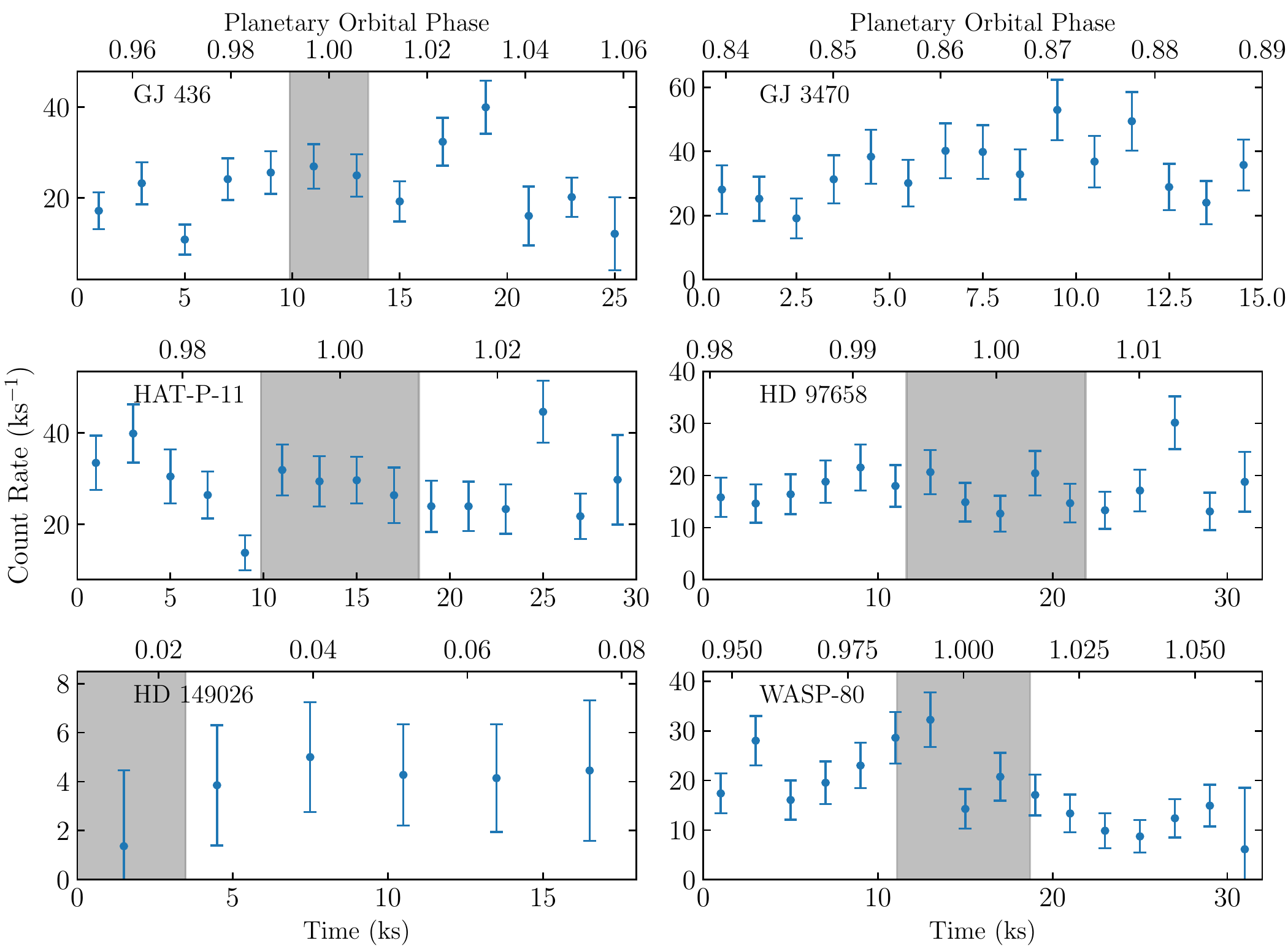}
 \caption{Background corrected X-ray light curves of the six targets. The count rate is the sum of the three EPIC detectors. The areas shaded in grey are the planetary transits (1st to 4th contact) in visible light. Time in each case is that elapsed from the beginning of the observation, as listed in Table~\ref{tab:Obs}.}
 \label{fig:lightcurves}
\end{figure*}

We analysed the time dependence of the targets for two primary purposes. Firstly, we check for strong stellar flares that could bias the measurements of the quiescent X-ray flux. 
Second, as shown in Table~\ref{tab:Obs}, four of our observations coincide with full planetary transits, and a fifth contains partial transit coverage. We examined our light curves for evidence of planetary transit features.

Figure~\ref{fig:lightcurves} displays the background corrected light curves, coadded across the three EPIC detectors. The count rate of HD\,149026 is too low to detect any variability. Of the other five observations, GJ\,436 and WASP-80 show temporal variability at the 3-$\sigma$ level when tested against a constant, equal to the mean count rate. HAT-P-11 also experiences variation, with a significance just below 3-$\sigma$. However, no strong flares are detected in any of the data, and none of the five observations covering a transit show any evidence of transit features in their light curves at this precision.

\subsection{X-ray spectra}
\label{ssec:Xspectra}

\begin{figure*}
\centering
 \includegraphics[width=\textwidth]{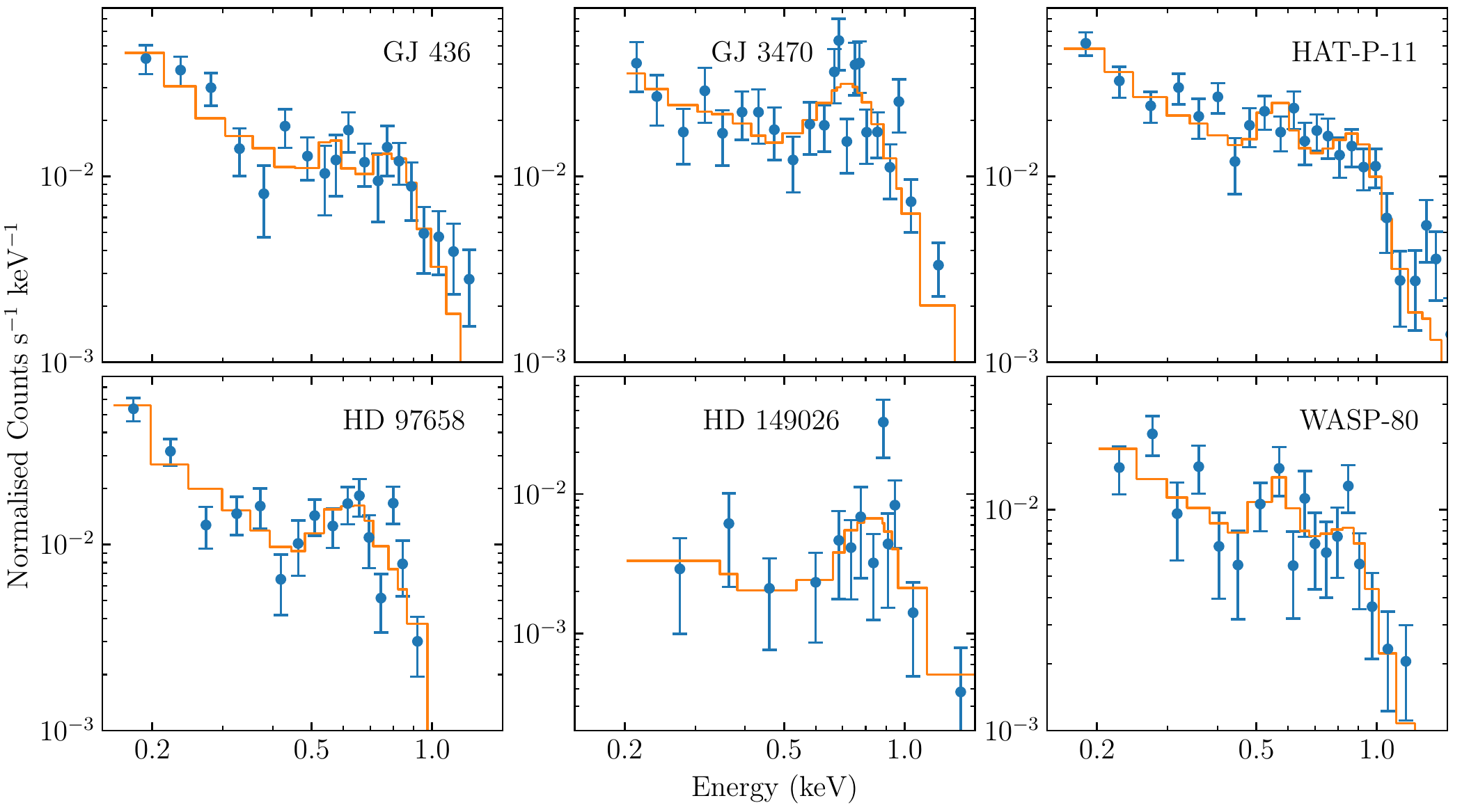}
 \caption{EPIC-pn X-ray spectra for the six targets. Unlike in the main analysis, the spectra are binned to a lower resolution to aid inspection. The background-corrected count rates are shown by the points with errorbars, with the histogram representing the fitted two-temperature APEC model.}
 \label{fig:spectra}
\end{figure*}

We analysed the unbinned, background corrected spectra in \textsc{xspec} 12.9.0~\citep{Arnaud1996}. Accordingly, we used C-statistics in our subsequent model fitting~\citep{Cash1979}. The errors on our fitted parameter values were determined using \textsc{xspec}'s error command, with confidence intervals of 68 per cent.

We fitted APEC models for optically-thin plasma in a state of collisional ionisation equilibrium~\citep{Smith2001}. In all cases with this model, a single-temperature fit can be rejected at 95 per cent confidence, using a Monte Carlo technique to assess the goodness of fit. We therefore performed fits with two temperature components, which gives a good fit for all six datasets. The abundances were fixed to solar values~\citep{Asplund2009}. Additionally, we included in a term for the interstellar absorption, making use of the TBABS model~\citep{Wilms2000}. We set the \ion{H}{I} column density for GJ\,436 and HD\,97658 to the values found by~\citet{Youngblood2016}. For the other four objects 
we follow the approach of~\citet{Salz2015}, who fixed the \ion{H}{I} column density to the distance of the system multiplied by a mean interstellar hydrogen density of 0.1\,cm$^{-3}$~\citep{Redfield2000}. We note that this estimate applies only to the Local Interstellar Cloud, and is not strictly applicable for lines of sight that contain other interstellar clouds. \citet{Redfield2008} showed that lines of sight to nearby stars varied around the average $N_{\rm H}$ value by about a factor of three. We found that changing $N_{\rm H}$ by a factor of three in either direction only changes the best fit measured fluxes by a few percent, well within the measured uncertainties.

The APEC-fitted EPIC-pn X-ray spectra are shown in Fig.~\ref{fig:spectra}. These have been binned to lower resolution to aid visualisation. The X-ray fluxes at Earth for the directly observed 0.2 -- 2.4\,keV band, $F_{\text{X, $\oplus$}}$, are shown in Table~\ref{tab:Results}. To obtain the unabsorbed fluxes, we changed the \ion{H}{I} column density to zero on the fitted model and reran the flux command. Since the error command cannot be run without refitting the model, we scaled the uncertainties so as to keep the percentage error constant between the absorbed and unabsorbed fluxes.

\subsection{X-ray fluxes}
\label{ssec:Xfluxes}

Most commonly used energy ranges for X-ray fluxes in the literature are conventions resulting from the passbands of various observatories. The ROSAT band (0.1 -- 2.4\,keV; 5.17 -- 124\,\AA) is one of the most widely employed. However, this band is not so easily applied to data from the current generation of X-ray observatories: the effective area of \textit{XMM-Newton's} EPIC pn and \textit{Chandra's} ACIS-S both decline quickly below 0.25\,keV. Thus, extrapolations to the ROSAT energy range must be made. As highlighted in~\citet{Bourrier2017} and~\citet{Wheatley2017}, while the fluxes obtained in \textsc{xspec} are usually seen to be consistent with one another in directly observed bands, the fluxes when extrapolating down to 0.1 keV can disagree significantly between models. 

We investigated the extrapolation discrepancy between the APEC model and CEMEKEL, the second model used by~\citet{Bourrier2017} and~\citet{Wheatley2017}. The latter is a multi-temperature plasma emission model, wherein the emission measure as a function of temperature is described by a power law~\citep{Schmitt1990,Singh1996}. The TBABS term accounting for interstellar absorption was applied in the same way as for the APEC model, above. For HD\,97658, the CEMEKL model yields a flux in the 0.2 -- 2.4\,keV band of $\left(2.90^{+0.15}_{-0.24}\right)\times10^{-14}$\flux, in good agreement with the APEC value of $\left(2.92^{+0.16}_{-0.77}\right)\times10^{-14}$\flux. However, when extrapolated down to 0.1\,keV, the CEMEKL value, 
$1.34\times10^{-13}$\flux, was almost four times lower than the corresponding APEC value of 
$5.3\times10^{-13}$\flux. Similar, but smaller, differences were also observed for the other objects. These originate from the different temperature-emission measure distribution assumptions of the models. For this reason, we chose to extrapolate directly from the observed X-ray fluxes to the full XUV band, rather than taking a two-step method of extrapolating to the \textit{ROSAT} band and then the EUV.

\begin{table*}
	\centering
	\caption{Results from our X-ray and EUV reconstruction analyses. The results given are for the X-ray range 0.2 -- 2.4\,keV, and corresponding EUV range 0.0136 -- 0.2\,keV. The X-ray fluxes at Earth are the APEC modelled values.}
	\label{tab:Results}
	\hspace*{-1.0cm}
	\begin{threeparttable}
	\begin{tabular}{lccccccccccc} 
		\hline
		System & $kT$ & $N_{\rm H}$ & EM & $F_{\text{X, $\oplus$}}$ & $L_\text{X}$ & $L_{\text{EUV}}$ & $F_{\text{XUV, p}}$ & $F_{\text{XUV, 1\,au}}$ & $L_{\text{Ly}\alpha}^\dagger$ & $F_{\text{Ly}\alpha, \oplus}^\dagger$ & $F_{\text{Ly}\alpha, \oplus}$ \\
		 & keV & (\textit{a}) & (\textit{b}) & (\textit{c}) & (\textit{d}) & (\textit{d}) & (\textit{e}) & (\textit{e}) & (\textit{d}) & (\textit{c}) & (\textit{c}) \\
		\hline
		GJ\,436 & \begin{tabular}[c]{@{}c@{}}$0.12\pm0.01$\\$0.61\pm0.08$\end{tabular} & 1.1 &
				\begin{tabular}[c]{@{}c@{}}$0.19\pm0.3$\\ $0.048^{+0.007}_{-0.006}$\end{tabular}
				 & $2.91^{+0.16}_{-0.27}$ & $0.33^{+0.02}_{-0.03}$ & $3.0^{+0.2}_{-0.3}$ &
				 $1380^{+100}_{-150}$ & $1.16^{+0.12}_{-0.16}$ & $3$ & $27$ & $20^\ast, 21^\S$ \\[0.3cm]
		GJ\,3470 & \begin{tabular}[c]{@{}c@{}}$0.09\pm0.03$\\ $0.35^{+0.06}_{-0.03}$\end{tabular} & 8.9 					 & \begin{tabular}[c]{@{}c@{}}$1.7^{+4.1}_{-0.9}$\\ $1.0^{+0.2}_{-0.2}$\end{tabular}
				 & $4.5^{+0.2}_{-0.9}$ & $4.5^{+0.8}_{-1.2}$ & $14.3^{+3.7}_{-4.9}$ & $4900^{+1000}_{-1300}$ &
				 $6.7^{+1.4}_{-1.8}$ & $13$ & $13$ & $-$ \\[0.3cm]
		HAT-P-11 & \begin{tabular}[c]{@{}c@{}}$0.16\pm0.01$\\ $0.81^{+0.12}_{-0.06}$\end{tabular} & 12 &
				\begin{tabular}[c]{@{}c@{}}$2.7\pm0.2$\\ $0.91^{+0.11}_{-0.10}$\end{tabular}
				 & $3.58^{+0.17}_{-0.21}$ & $6.2^{+0.3}_{-0.4}$ & $22.2^{+2.5}_{-2.7}$ &
				 $3560^{+330}_{-350}$ & $10.07^{+0.95}_{-1.04}$ & $36$ & $21$ & $-$ \\[0.3cm]
		HD\,97658 & \begin{tabular}[c]{@{}c@{}}$0.044^{+0.013}_{-0.008}$\\ $0.24\pm0.1$\end{tabular} & 						2.8 & \begin{tabular}[c]{@{}c@{}}$15^{+16}_{-10}$\\ $0.56\pm{0.06}$\end{tabular}
				 & $2.92^{+0.16}_{-0.78}$ & $1.62^{+0.09}_{-0.43}$ & $11.1^{+1.1}_{-3.3}$ &
				 $700^{+60}_{-190}$ & $4.52^{+0.42}_{-1.25}$ & $22$ & $39$ & $42^\ddagger, 91^\S$ \\[0.3cm]
		HD\,149026 & \begin{tabular}[c]{@{}c@{}}$0.09^{+0.41}_{-0.06}$\\ $0.71^{+0.14}_{-0.09}$ 							\end{tabular} & 24 & \begin{tabular}[c]{@{}c@{}}$2.3^{+148.6}_{-2.3}$\\$1.7^{+0.4}_{-0.2}$\end{tabular} & $0.84^{+0.01}_{-0.21}$ & $5.9^{+0.2}_{-1.5}$ & $37^{+5}_{-11}$ & $8300^{+900}_{-2100}$ &
				 $15.2^{+1.7}_{-3.9}$ & $52$ & $7.4$ & $-$ \\[0.3cm]
		WASP-80 & \begin{tabular}[c]{@{}c@{}}$0.15\pm0.02$\\ $0.73\pm0.09$ \end{tabular} & 19 &
				\begin{tabular}[c]{@{}c@{}}$3.6^{+2.8}_{-2.0}$\\ $1.1^{+0.9}_{-0.6}$\end{tabular}
				 & $1.78^{+0.11}_{-0.16}$ & $8\pm5$ & $19\pm14$ & $8900\pm4300$ &  $9.5\pm5.3$ & $31$ &
				 $7.3$ & $-$ \\
		\hline
	\end{tabular}
    \begin{tablenotes}
\item \textit{a} $10^{18}$\,cm$^{-2}$ (column density of H)
\item \textit{b} $10^{50}$\,cm$^{-3}$
\item \textit{c} $10^{-14}$\flux\ (at Earth, unabsorbed)
\item \textit{d} $10^{27}$\,erg\,s$^{-1}$
\item \textit{e} \flux\
\item $^\dagger$ Estimated using the relations between EUV and Ly\,$\alpha$ fluxes at 1\,au in~\citet{Linsky2014}.
\item $^\ast$ As reconstructed from observation by~\citet{Bourrier2016}.
\item $^\ddagger$ As reconstructed from observation by~\citet{Bourrier2017}.
\item $^\S$ As reconstructed by~\citep{Youngblood2016}.
    \end{tablenotes}
  \end{threeparttable}
\end{table*}

\subsection{EUV reconstruction}
\label{ssec:EUVrec}

EUV fluxes of stars must be reconstructed using other spectral ranges. \citet{Salz2015} compared three such methods, finding them to differ by up to an order of magnitude in active stars. However, \citet{Chadney2015}, hereafter C15, presented a new empirical method of reconstructing the EUV flux from the measured X-ray flux. This method shows a better agreement with stellar rotation-based and stellar Ly\,$\alpha$ luminosity-based reconstructions~\citep{LDE2007,Linsky2014} than the X-ray-based method of \citet{SanzForcada2011}.

C15 analysed observations of the Sun, deriving a power law relation between the ratio of EUV to X-ray flux and the X-ray flux. This method seems physically well motivated, relating the fluxes at the stellar surface, thereby implicitly taking the local conditions of this region into account. Indeed, their result agrees well with synthetic spectra for a small number of nearby, K and M dwarf stars, as generated from coronal models. These synthetic spectra, in turn, agree with EUVE measurements within the uncertainties.

The C15 relation adopts the ROSAT band. Accordingly, they define the EUV band as 0.0136 -- 0.1 keV (124 -- 912 \AA). As discussed in section~\ref{ssec:Xfluxes}, this definition does not transfer well onto the current generation of X-ray telescopes. To apply the C15 relation to observations by either \textit{XMM-Newton} or \textit{Chandra}, one must perform two extrapolations. The first estimates the missing X-ray flux down to 0.1 keV, which we have shown to be uncertain by a factor of a few. The second occurs in applying the relation itself. Given the model-dependence on the first of these steps highlighted above, it would be preferable to derive a new set of relations that allow direct extrapolation from the observed band to the rest of the XUV range in a single step. By reperforming the C15 analysis with different boundaries, we can derive such new relations that are more applicable to current instruments.

\subsubsection{Derivation of new X-ray-EUV relations}
\label{sssec:NewRel}

The data used by C15 comes from the ongoing \textit{TIMED/SEE} mission~\citep{Woods2005}. One of the primary data outputs of the mission is daily averaged Solar irradiances, given in 10\,\AA\ intervals from 5 -- 1945\,\AA. We integrated the fluxes up to the Lyman limit (0.0136\,eV, 912\,\AA), splitting the data into X-ray and EUV bands either side of some defined boundary. Here, we used a range of boundary choices to produce our set of relations.

\begin{figure}
 \includegraphics[width=\columnwidth]{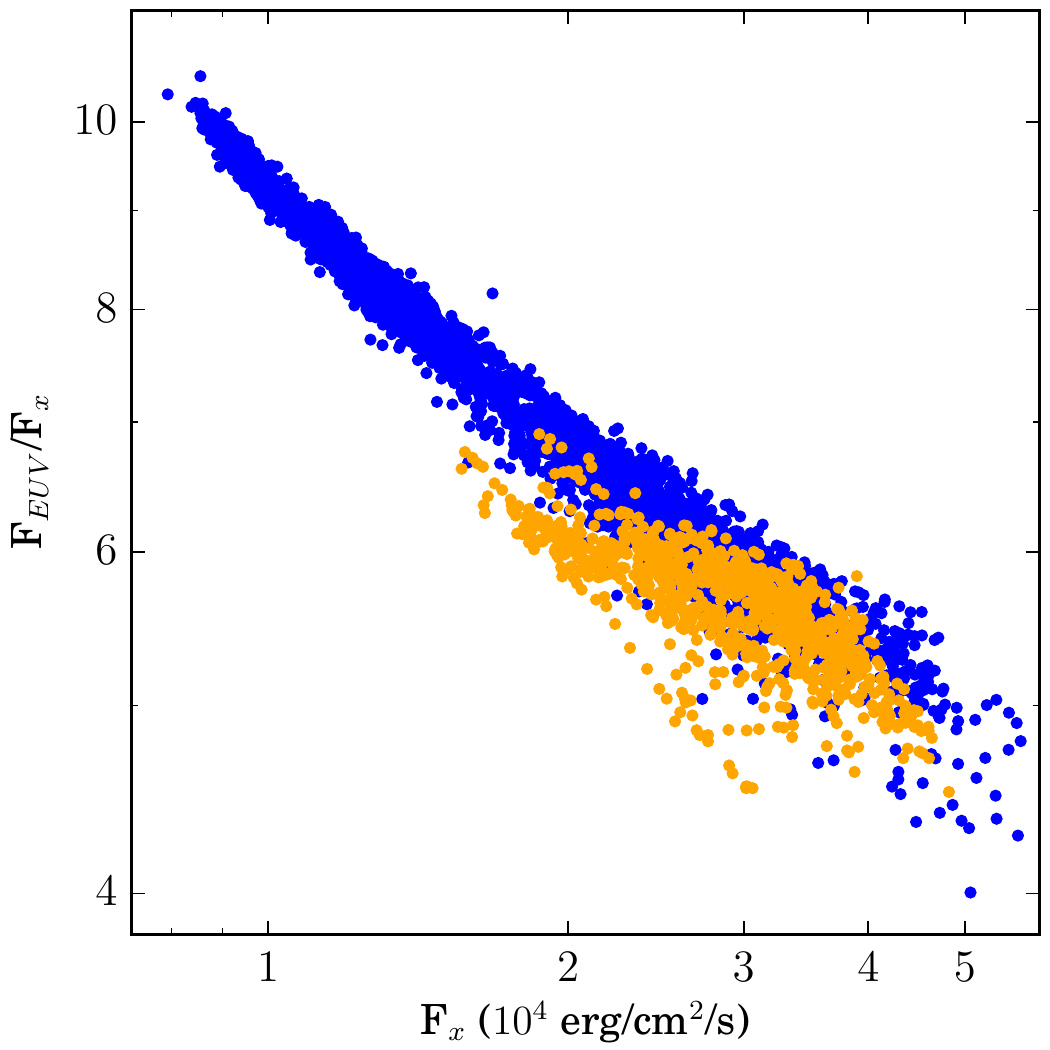}
 \caption{An updated version of fig.~2 of C15: the ratio of the EUV flux to X-ray flux plotted against the X-ray flux for a boundary energy of 0.1 keV. The C15 sample (30 May 2002 -- 16 November 2013) is shown in blue, and data from 17 November 2013 to 21 July 2016 are shown in orange.}
 \label{fig:C15fig2}
\end{figure}

\begin{figure}
 \includegraphics[width=\columnwidth]{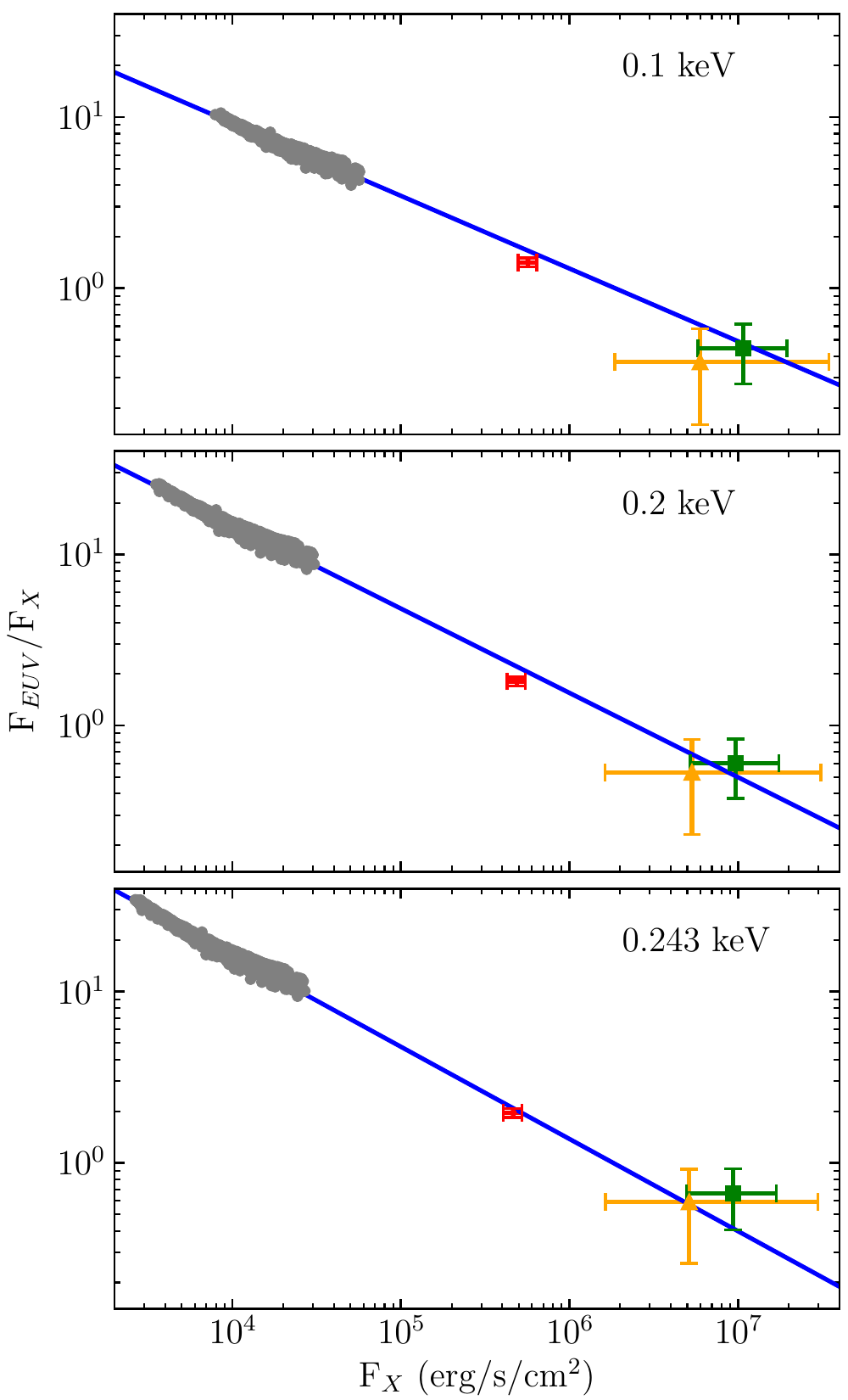}
 \caption{Solar \textit{TIMED/SEE} data plotted for three of the new boundary energy choices: 0.1\,keV/124\,\AA\ (top), 0.2\,keV/62\,\AA\ (middle), and 0.243\,keV/51\,\AA\ (bottom). Fluxes for the comparison stars are plotted as follows: $\epsilon$\,Eri - red circle; AD\,Leo - orange triangle; AU\,Mic - green square.}
 \label{fig:NewRelations}
\end{figure}

Using only the C15 sample (30 May 2002 -- 16 November 2013), we were able to replicate their relation exactly, but we have the benefit of extra data. However, we noticed that some of the most recent observations appear to be offset from the rest of the data (see Fig.~\ref{fig:C15fig2}). This offset is likely a result of instrument degradation, which is not yet properly accounted for in the recent data (private communication with the TIMED/SEE team). Therefore, we chose to cut off all data past 1 July 2014, where the data start to show significant differences to older observations.
 
Additionally, we noticed that the errorbars in the merged file of all observations did not match those in the individual daily files which was kindly fixed by the mission team. It seems that the data used by C15 had the same problem, so we also update C15's relation for the 0.1 keV boundary.

Fig.~\ref{fig:NewRelations} shows the solar \textit{TIMED/SEE} data and fluxes from the comparison synthetic stellar spectra, plotted for three of the boundary choices. The residuals of the single power law fit reveal a trend. As the choice of boundary energy is increased, the log-log plot increasingly deviates from linear. A more complex function may be justified when solely considering the solar data. However, this would have proved less robust when extrapolating the relation to higher flux levels in active stars. We obtained synthetic spectra for a sample of nearby stars: $\epsilon$\,Eri from the X-Exoplanets archive\footnote{Available at~\url{http://sdc.cab.inta-csic.es/xexoplanets/jsp/homepage.jsp}. See also~\citet{SanzForcada2011}.}, and the spectra for AD\,Leo and AU\,Mic presented in C15. Using these, a single power law fitted to the solar data agrees well with the comparison stars. During this comparison process, we also found that unweighting the solar data actually provided a slightly better fit with regard to the comparison stars across the choice of boundary energies.

Given the choice of a single power law, each relation takes the form
\begin{equation}
\frac{F_{\text{EUV}}}{F_{\text{X}}} = \alpha\left(F_{\text{X}}\right)^\gamma,
\label{eq:relation}
\end{equation}
where $F_{\text{EUV}}$ is the flux in the extrapolated band, from 0.0136\,keV up to the chosen boundary, and $F_{\text{X}}$ is the flux in the observed band, from the boundary up to 2.4\,keV. The exception to this is the 0.124\,keV boundary which, as per convention, extends the observed band to 2.48\,keV. As in C15, these fluxes are those at the stellar surface. The values of $\alpha$ and $\gamma$ are given in Table~\ref{tab:RelVar} for each of the five boundary choices. As highlighted in Table~\ref{tab:RelVar}, each of the boundary energies were chosen to correspond to the observational band of an X-ray satellite, or a widely-used choice in the literature. We also include two further relations for going directly from the 0.2 -- 2.4\,keV band to the 0.0136 -- 0.1\,keV and 0.0136 -- 0.124\,keV EUV bands.

\begin{table*}
	\centering
	\caption{Best fitting power laws to be used in conjunction with equation~\ref{eq:relation} for the each choice of boundary energy. See Sect.~\ref{sssec:NewRel}.}
	\label{tab:RelVar}
	\begin{tabular}{cccccccc} 
		\hline
		\# & \multicolumn{2}{c}{X-ray range} & \multicolumn{2}{c}{EUV range} & $\alpha$ & $\gamma$ & Relevant Satellite \\
		 & (keV) & (\AA) & (keV) & (\AA) & (\flux) &  &   \\
		\hline
		1 & 0.100 -- 2.400 & 5.17 -- 124 & 0.0136 -- 0.100 & 124 -- 912 & 460 & -0.425 & \textit{ROSAT} (PSPC) \\
		2 & 0.124 -- 2.480 & 5.00 -- 100 & 0.0136 -- 0.124 & 100 -- 912 & 650 & -0.450 & None, widely-used (5 -- 100\,$\AA$) \\
		3 & 0.150 -- 2.400 & 5.17 -- 83 & 0.0136 -- 0.150 & 83 -- 912 & 880 & -0.467 & \textit{XMM-Newton} (pn, lowest) \\
		4 & 0.200 -- 2.400 & 5.17 -- 62 & 0.0136 -- 0.200 & 62 -- 912 & 1400 & -0.493 & $\left\{
	 \begin{tabular}[c]{@{}c@{}}\textit{XMM-Newton} (pn, this work),\\
	 \textit{XMM-Newton} (MOS),\\
	 \textit{Swift} (XRT)\end{tabular} \right. $\\
		5 & 0.243 -- 2.400 & 5.17 -- 51 & 0.0136 -- 0.243 & 51 -- 912 & 2350 & -0.539 & \textit{Chandra} (ACIS)\\
		\hline
		6 & 0.200 -- 2.400 & 5.17 -- 62 & 0.0136 -- 0.100 & 124 -- 912 & 1520 & -0.509 & \textit{XMM-Newton} (Observed to ROSAT EUV)\\
		7 & 0.200 -- 2.400 & 5.17 -- 62 & 0.0136 -- 0.124 & 100 -- 912 & 1522 & -0.508 & \textit{XMM-Newton} (Observed to 5 -- 100\,$\AA$ band)\\
		\hline
 		\end{tabular}
\end{table*}

\subsubsection{Total XUV flux calculations}
\label{sssec:XUVcalc}
Using our newly derived relations, we determine the full XUV flux at the stellar surface, at the distance of each planet, $F_{\text{XUV}, p}$, and at 1\,au (see Table~\ref{tab:Results}). For the zero eccentricity planets GJ\,3470b and HD\,149026b, we simply use the semi-major axis in Table~\ref{tab:SysParam}. WASP-80b has a small upper limit on its eccentricity, so we again use the semi-major axis estimate. However, GJ\,436, HAT-P-11, and HD\,97658 all have non-zero eccentricities, and as such we use the time-averaged separation~\citep[see, for a discussion,][]{Williams2003}. Consequently, determined values of $F_{\text{XUV}, p}$ in these cases should also be considered time-averages. We find that HD\,149026b and WASP-80b are subject to the largest XUV irradiation. GJ\,3470b and HAT-P-11b receive about half the XUV flux of HD\,149026b and WASP-80b, but still a few times more than GJ\,436b and HD\,97658b. 

Note that the XUV luminosity, and so $F_{\text{XUV}, p}$, of WASP-80 are subject to larger uncertainty. This is a consequence of its poorly-known distance of $60\pm20$\,pc~\citep{Triaud2013}. No parallax was given in first \textit{Gaia} data release, even though the star has a Tycho designation.

\section{Optical Monitor results}
\label{sec:OMres}
Observations using the OM camera on \textit{XMM-Newton} were taken concurrently with those of the EPIC X-ray detectors. Different observing strategies were employed for this instrument in the two separate proposals that comprised the full set of observations we describe. In both cases, however, we have taken advantage of the near NUV capabilities of the OM.

\subsection{GJ\,3470 and HD\,149026}

\begin{table}
	\centering
	\caption{OM results for GJ\,3470 and HD\,149026.}
	\label{tab:OMSalz}
	\begin{threeparttable}
	\begin{tabular}{lccc} 
		\hline
		Filter & Central $\lambda$ & Flux & Mag. \\
		 & (\AA) & $10^{-15}$\flux\AA$^{-1}$ &  \\
		\hline
		\multicolumn{4}{c}{\textit{GJ\,3470}} \\[0.1cm]
		U & 3440 & $3.66\pm0.07$ & 14.9 \\
		UVW1 & 2910 & $0.49\pm0.24$ & 17.2 \\
		UVM2 & 2310 & $1.2^\ast$ & 16.4 \\
		\hline
		\multicolumn{4}{c}{\textit{HD\,149026}} \\[0.1cm]
		UVM2 & 2310 & $0.10\pm0.02$ & 19.1 \\
		UVW2 & 2120 & $0.30\pm0.04$ & 18.1 \\
		\hline
		\end{tabular}
    \begin{tablenotes}
\item $^\ast$ Note that the UVM2 flux conversion introduces a factor of two error for M dwarf stars.
    \end{tablenotes}
  \end{threeparttable}
\end{table}

For GJ\,3470 and HD\,149026, some of the ultraviolet filters were cycled through in turn during the observation period. In the case of GJ\,3470, all ultraviolet filters were employed except UVW2, that pushes furthest into the ultraviolet but is also the least sensitive. All ultraviolet filters were used for HD\,149026, but the object was saturated in the U and UVW1 filters, leaving useful measurements only for UVW2 and the next bluest ultraviolet filter, UVM2.

For both objects, the measured count rates were converted into fluxes and magnitudes following the prescription of a \textsc{sas} watchout page\footnote{``How can I convert from OM count rates to fluxes", available at \url{https://www.cosmos.esa.int/web/xmm-newton/sas-watchout-uvflux}.}. We adopted the conversions for M0V and G0V stars for GJ\,3470 and HD\,149026, respectively (cf. spectral types in Table~\ref{tab:SysParam}). The calculated fluxes and magnitudes for each filter used for each object are summarised in Table~\ref{tab:OMSalz}.

\subsection{Fast mode observations}
\label{ssec:fastmodeOM}

\begin{figure}
 \includegraphics[width=\columnwidth]{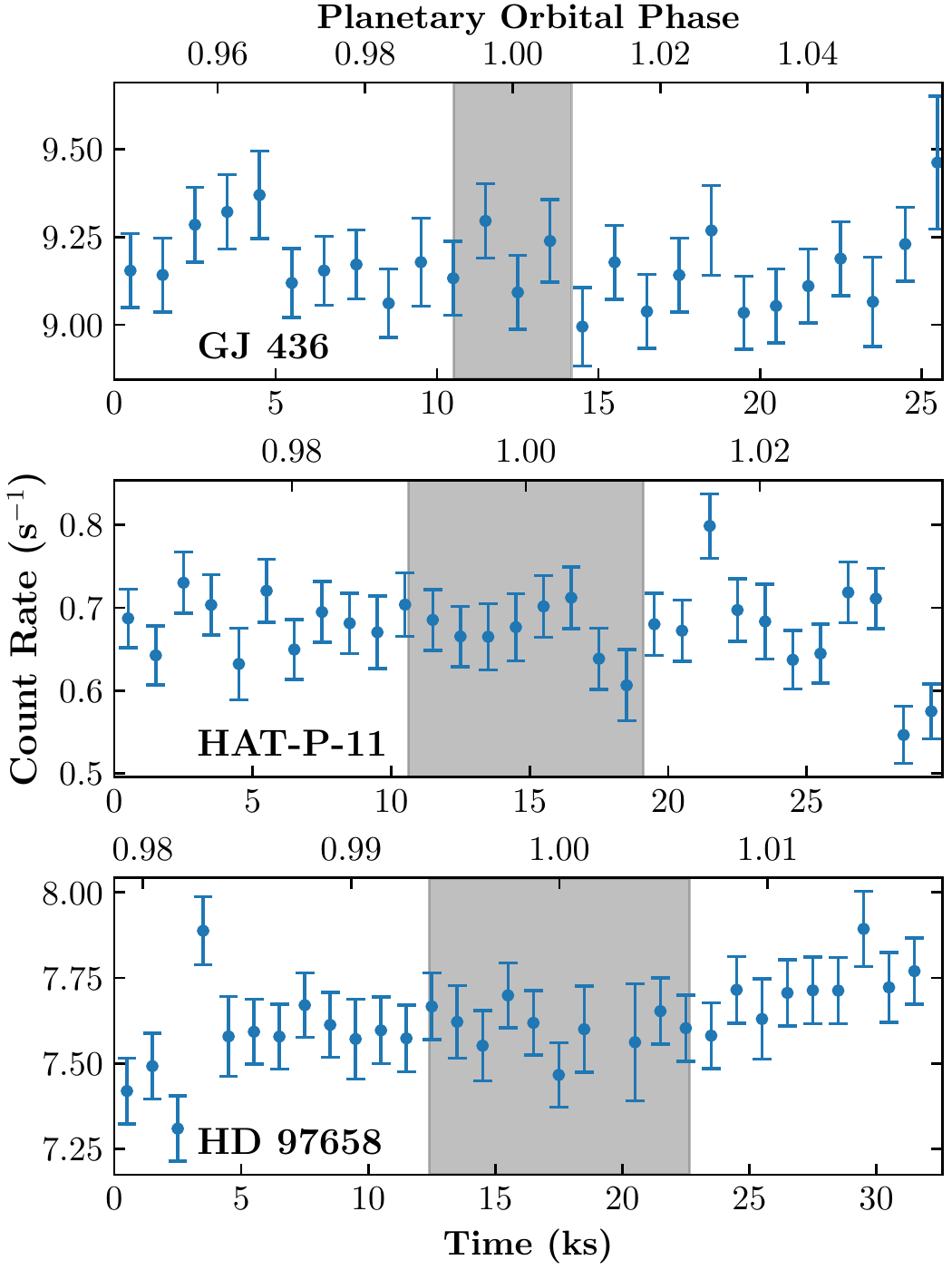}
 \caption{Optical Monitor light curves for GJ\,436, HAT-P-11, and HD\,97658, binned to 1000\,s resolution. The areas shaded in grey are the planetary transits (1st to 4th contact) in visible light.}
 \label{fig:om_comb}
\end{figure}

The other four objects were observed in a single filter, and in fast mode, in order to probe ultraviolet variation in the source over the course of the observation. This opened up the possibility of detecting the transit in the NUV. In each case, the single filter choice was a trade off between wishing to push as far into the NUV as possible, while wanting to maintain a high enough (predicted) count rate that transit detection level precision might be possible. UVW1 was chosen for GJ\,436 and WASP-80, while HD\,97658 and HAT-P-11 were observing using the UVW2 filter.

The final light curves for GJ\,436, HAT-P-11 and HD\,97658 are shown in Fig.~\ref{fig:om_comb}, and we conclude that none of these three observations detected the transit in NUV. The light curves were built by correcting the fast mode time series data from \textsc{omfchain} using the corresponding image mode extractions from \textsc{omichain}. The reasons for this are described in Appendix~\ref{Appen:OM}.

\subsubsection{WASP-80}
\label{sssec:WASP-80_OM}

We identified a possible transit detection in the WASP-80 data. Again, we correct the fast mode time series by the corresponding image mode extractions, as described in Appendix~\ref{Appen:OM}.

\begin{figure}
 \includegraphics[width=\columnwidth]{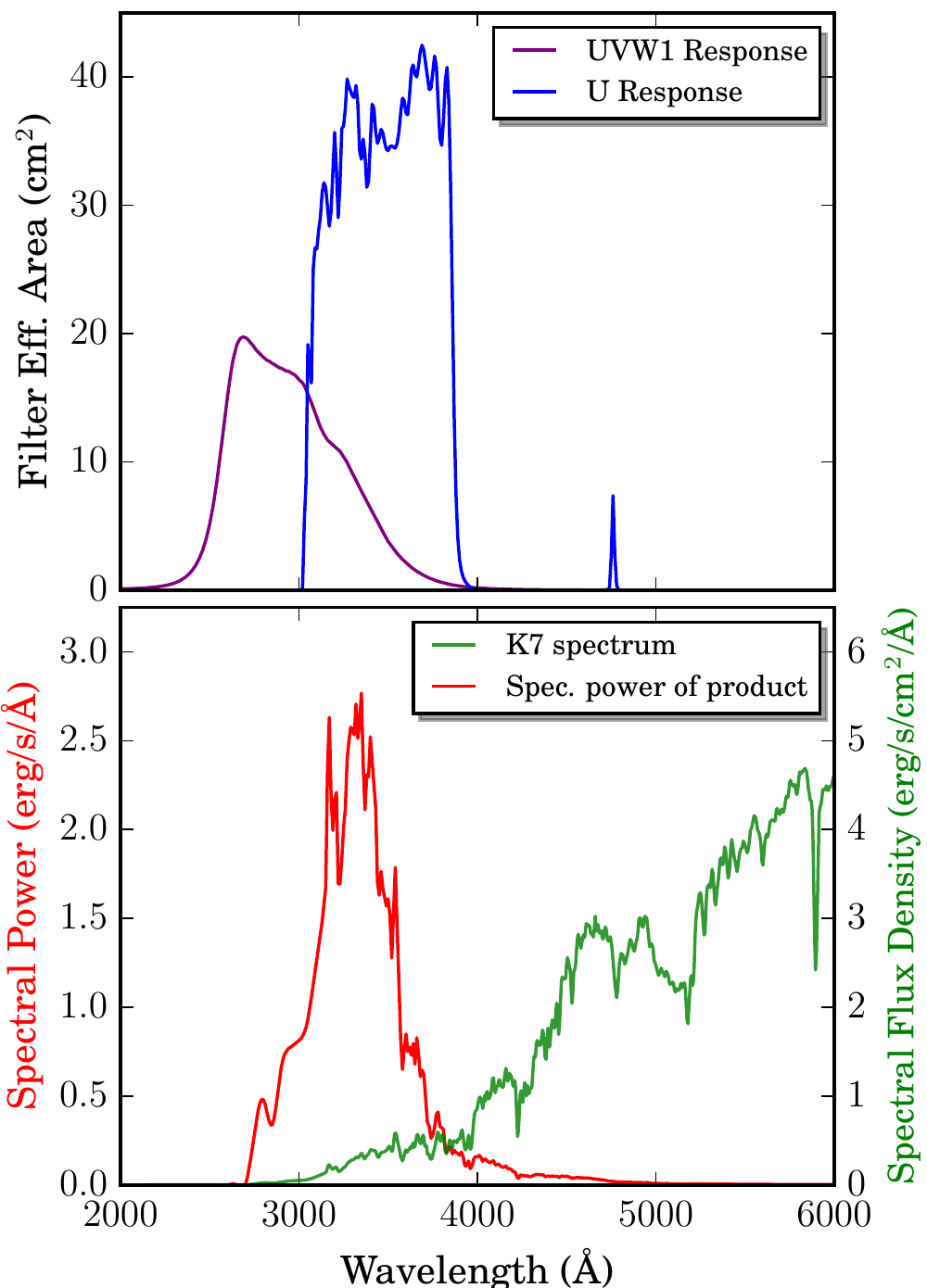}
 \caption{Top: Effective area of the UVW1 (purple) and U band filters on the OM camera as a function of wavelength. Bottom: Model spectrum for a K7V star, and the product of the UVW1 effective area and the K7V spectrum.}
 \label{fig:filtResp}
\end{figure}

\begin{figure}
 \includegraphics[width=\columnwidth]{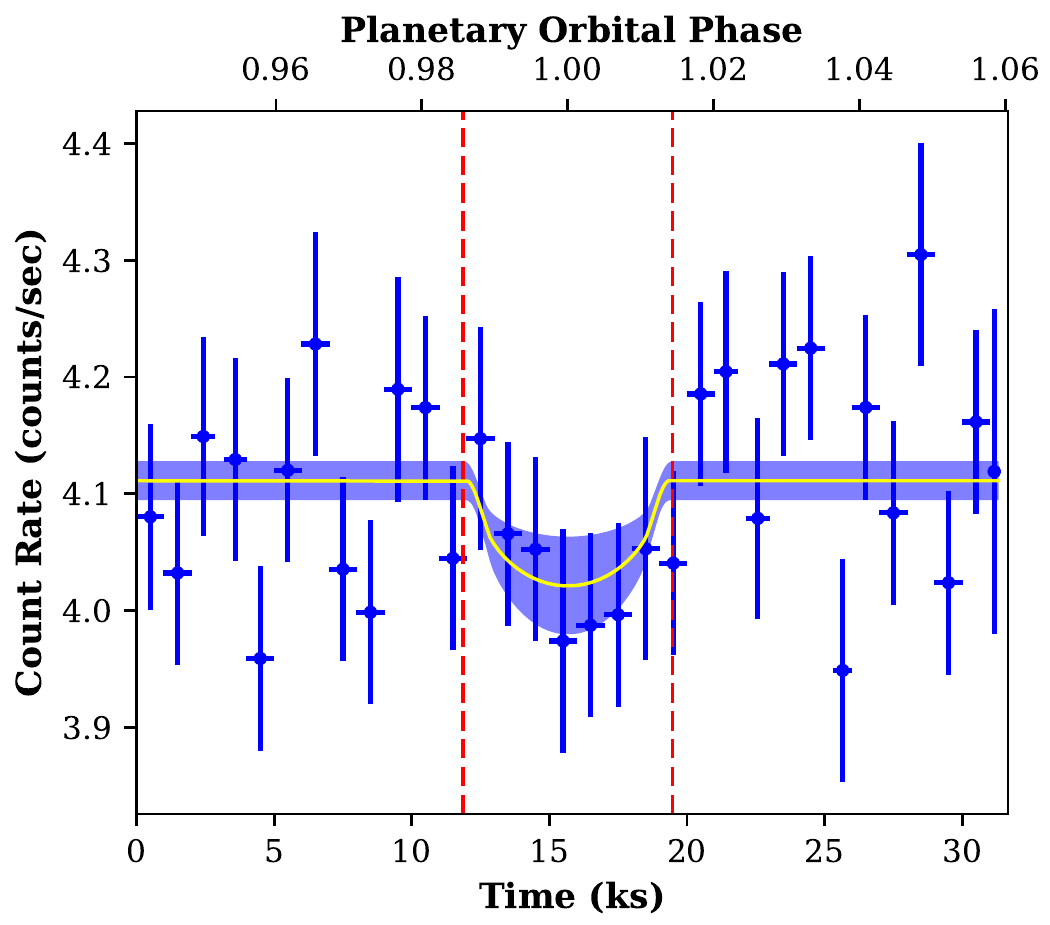}
 \caption{WASP-80 data binned to 1000\,s bins. Overlaid is the best fit model (yellow) along with the 1-$\sigma$ confidence region (blue shaded region). The dotted red lines correspond to the first and fourth contact of the transit, as calculated from the visible light ephemeris~\citep{Mancini2014}.}
 \label{fig:wasp80om}
\end{figure}

\begin{figure}
 \includegraphics[width=0.85\columnwidth]{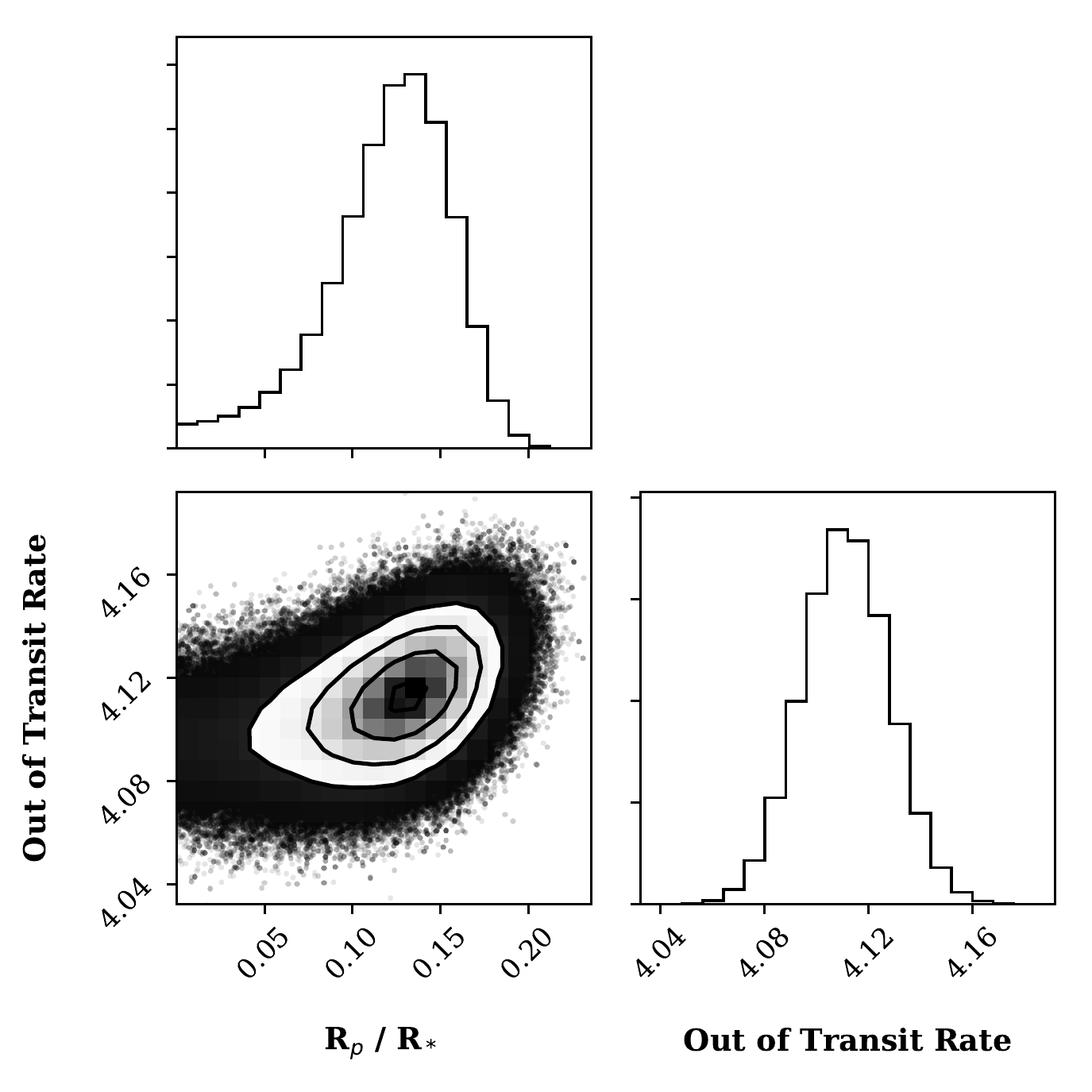}
 \caption{Corner plot for the WASP-80 fit showing the correlation of the out of transit count rate with $R_p / R_*$. The parameters bound by a Gaussian prior are omitted.} 
 \label{fig:wasp80corner}
\end{figure}

\begin{table}
	\centering
	\caption{WASP-80 near ultraviolet MCMC fit priors and results.}
	\footnotesize
	\label{tab:wasp80om}
	\begin{tabular}{lcc} 
		\hline
		Parameter & Value & Reference \\
		\hline
		\multicolumn{3}{c}{\textit{Gaussian priors}} \\[0.1cm]
		$t_{\text{Cen}}$ (BJD) & $2457156.21885(31)$ & \citet{Mancini2014} \\
		$a/R_*$ & $12.989\pm0.029$ & \citet{Triaud2013} \\
		$i$ & $89.92\pm0.10$ & \citet{Triaud2013} \\
		\hline
		\multicolumn{3}{c}{\textit{Fixed values}} \\[0.1cm]
		u1 & 0.9646 & \citet{Claret2011} \\
		u2 & $-0.1698$ & \citet{Claret2011} \\
		\hline
		\multicolumn{3}{c}{\textit{Free, fitted parameter}} \\[0.1cm]
		$R_p / R_*$ & $0.125^{+0.029}_{-0.039}$ & \textit{This work} \\
		\hline
		\end{tabular}
\end{table}

We modelled our time series using the \textsc{transit} code\footnote{Available as part of the \textsc{rainbow} package (\url{https://github.com/StuartLittlefair/rainbow}). Documentation can be found at \url{http://www.lpl.arizona.edu/~ianc/python/transit.html}.}, a \textsc{python} implementation of the~\citet{Mandel2002} analytic transit model. To fit the model we used the MCMC sampler provided by the \textsc{emcee} package~\citep{FM2013}. We set Gaussian priors on the transit centre time, $a/R_*$, and the system inclination, $i$, according to the values and references in Tables~\ref{tab:SysParam} and~\ref{tab:Obs}. The prior for the transit centre at the epoch of our observations, $t_{\text{Cen}}$, was calculated using the ephemeris of~\citet{Mancini2014}. $R_p / R_*$ and the out of transit count rate were allowed to vary freely with uniform priors. The latter was included to normalise the out of transit data to an intensity of unity. 

The limb darkening coefficients were fixed to those for the U band from~\citet{Claret2011}, according to the stellar properties of WASP-80. Despite being taken with the UVW1 filter, the sampling of the late-K dwarf spectrum is weighted to the U band, due to the red leak of the filter. This is shown in Fig.~\ref{fig:filtResp} which plots the effective area of the OM UVW1 and U band filters, a model spectrum for a K7 dwarf star~\citep{Pickles1998}, and the product of the UVW1 response with the model spectrum. 

Fig.~\ref{fig:wasp80om} displays the WASP-80 OM light curve with the best fit model and the 1-$\sigma$ credibility region, with the data binned to a lower resolution to aid the eye. The resulting best fit parameters for the model are given in Table~\ref{tab:wasp80om}. The best fitting depth is shallower than previous optical measurements, but is consistent to within 1.6-$\sigma$. Our best fit $R_p/R_*$ shows some weak correlation with the out of transit count rate. The associated corner plot, made using the corner.py code~\citep{FM2016}, is shown in Fig.~\ref{fig:wasp80corner}.

\section{Discussion}
\label{sec:Discuss}

\subsection{X-ray Fluxes}
\label{ssec:X-rayEvo}
The link between coronal X-ray emission and rotation period has been explored extensively~\citep[e.g.][]{Pallavicini1981,Pizzolato2003,Wright2011,Jackson2012,Wright2016,Stelzer2016}. At the shortest rotation periods, i.e. early in a star's life, the X-ray emission is close to saturation, where the ratio of the X-ray emission to the bolometric luminosity, $L_\text{X}/L_\text{bol}$, is about $10^{-3}$. The rotation period, $P_\text{rot}$, of a star slows down as it ages. Once the rotation slows to beyond some critical value, $L_\text{X}/L_\text{bol}$, is seen to drop off with a power law behaviour.

\citet{Pizzolato2003} derived empirical relations describing $L_\text{X}$ and $L_\text{X}/L_\text{bol}$ as a function of $P_\text{rot}$. Further, they confirmed a relationship with Rossby number, $R_\text{o}$, for late-type stars with different convection properties. $R_\text{o}$ is defined as the ratio of $P_\text{rot}$ and $\tau$, the convective turnover time~\citep{Noyes1984}. \citet{Wright2011}, hereafter W11, formulated a set of empirical relations for this link between $L_\text{X}/L_\text{bol}$ and $R_\text{o}$. This alternative formulation of the relationship reduced the scatter among unsaturated stars. W11 also better constrains M stars due to its larger sample of such stars. This is useful for our study, which contains two M stars and a third on the K-M type boundary. \citet{Wright2016} further explored the application of this relation to low mass, fully convective stars with their observed $L_\text{X}/L_\text{bol}$ correlating well with the W11 relations. We compare our measured fluxes to the W11 relations.

The X-ray emission considered in W11 is for the 0.1 -- 2.4\,keV ROSAT band. Examining the solar TIMED/SEE data in a similar way to the method in Section~\ref{sssec:NewRel} with the two bands defined as 0.1 -- 0.2 and 0.2 -- 2.4\,keV showed an approximate 1:1 ratio of flux in the two bands. We therefore doubled the flux in the observed 0.2 -- 2.4\,keV to estimate that in the ROSAT band. However, we added 50 per cent uncertainties in quadrature with the observed flux errors, due to the scatter of the comparison stars to the TIMED/SEE data. $L_\text{bol}$ was evaluated using the Stefan-Boltzmann law. 
We note that the subgiant nature of HD\,149026 means that the W11 relations, derived for main sequence stars, may not be directly applicable to the star.

\begin{figure}
 \includegraphics[width=\columnwidth]{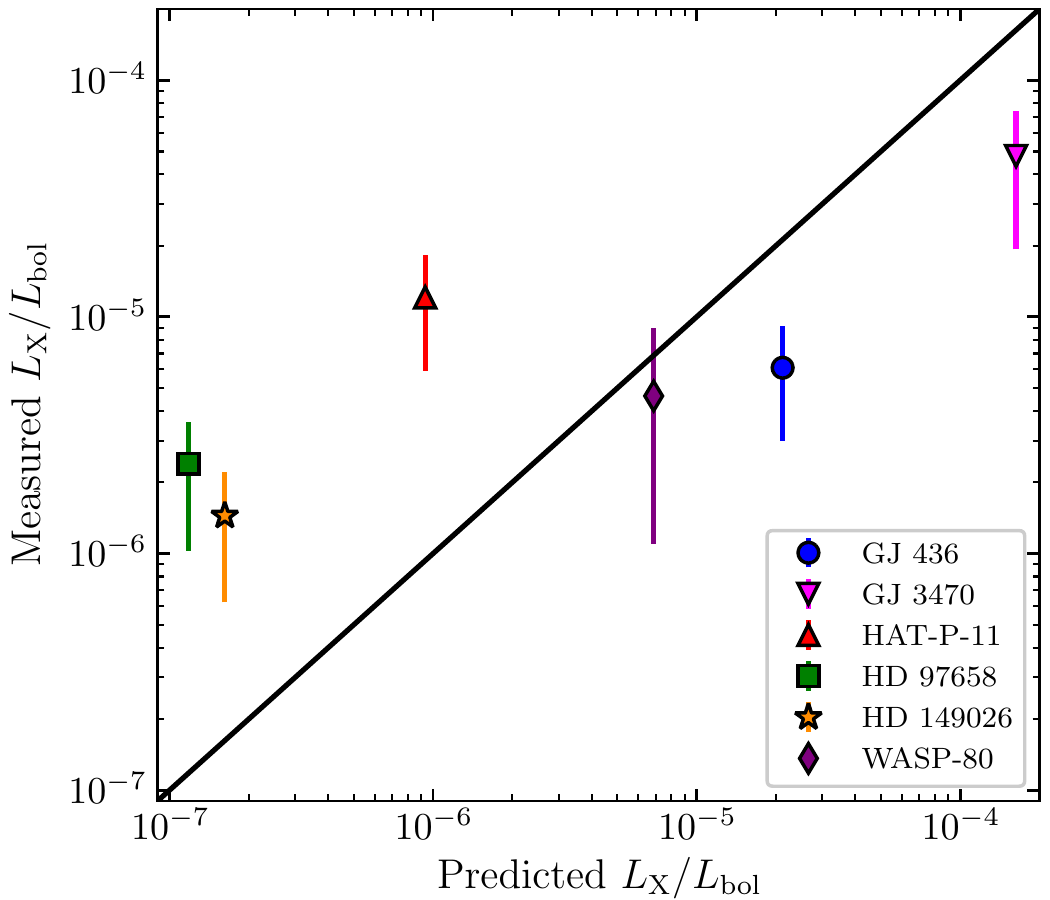}
 \caption{Comparison of the measured $L_\text{X}/L_\text{bol}$ to that expected from the relations of W11.}
 \label{fig:measVexpt}
\end{figure}

\begin{figure}
 \includegraphics[width=\columnwidth]{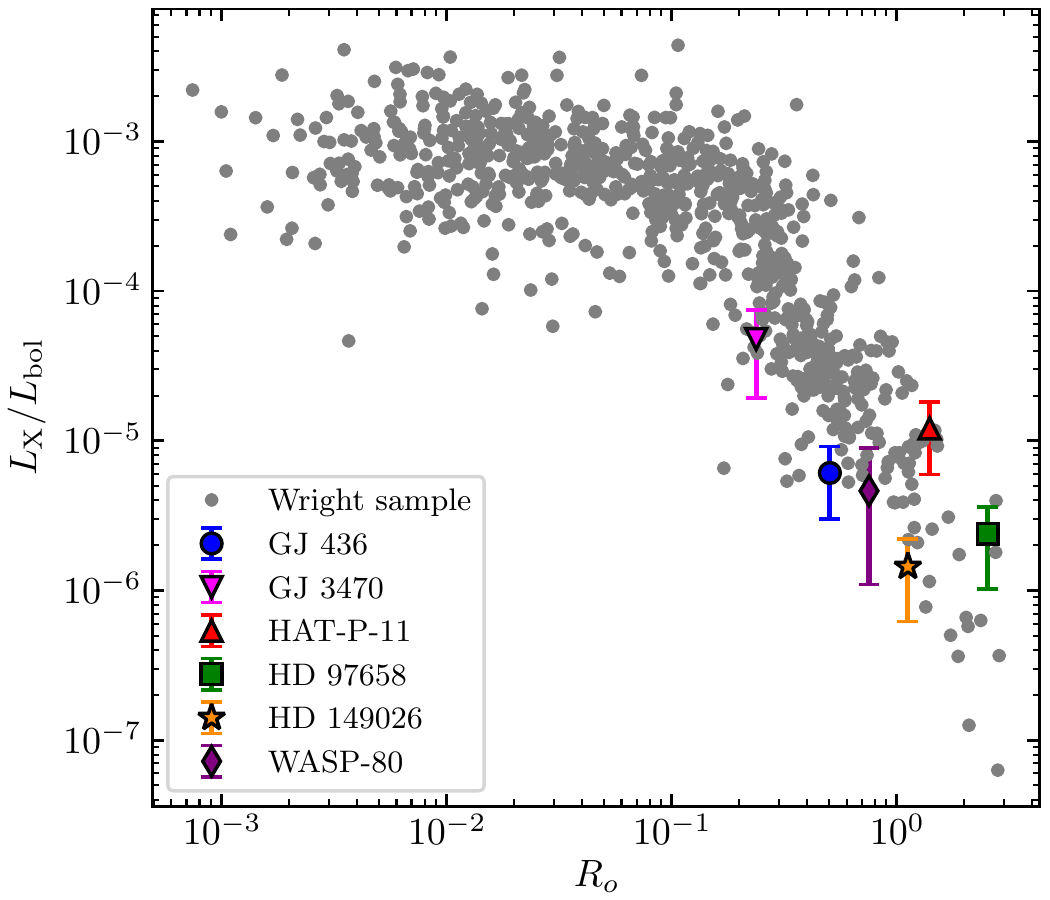}
 \caption{Replotting of fig.~1 of~\citet{Wright2016}, itself an update of the W11 sample, with points added from our own sample.}
 \label{fig:w16replot}
\end{figure}

Fig.~\ref{fig:measVexpt} depicts our measured $L_\text{X}/L_\text{bol}$ against that expected from W11. We note that our sample has a trend with slow rotators being more X-ray luminous than predictions, suggesting their activity may not drop as quickly as predicted. \citet{Booth2017} recently found a steeper age-activity slope for old, cool stars to previous studies. They suggested that in the context of the findings of~\citet{vanSaders2016}, which found evidence for weaker magnetic breaking in field stars older than 1\,Gyr, this could point to a steepening of the rotation-activity relationship, in contrast to our measurements. Despite the apparent shallower trend in Fig.~\ref{fig:measVexpt}, our measurements are in line with the scatter in the W11 sample itself, as can be seen in Fig.~\ref{fig:w16replot}. The significant scatter in these activity relations underlines the need for measurements of X-ray fluxes for individual exoplanet hosts.

\begin{table}
	\centering
	\caption{Comparison of GJ\,436 and WASP-80 X-ray fluxes with previous studies, grouped by energy range.}
	\footnotesize
	\label{tab:compFluxes}
	\hspace*{-0.25cm}
	\begin{threeparttable}
	\begin{tabular}{lccc} 
		\hline
		Dataset	& Reference & Energy Range	& Flux \\
				& 			& (keV)	 		& (\textit{a}) \\
		\hline
		\multicolumn{4}{c}{\textit{GJ\,436}} \\[0.1cm]
		2008, \textit{XMM}				& SF11		& 0.124 -- 2.48	& $0.73$ \\
		2008, \textit{XMM}				& E15		& 0.124 -- 2.48	& $4.6$ \\[0.25cm]
		2008, \textit{XMM}				& This work	& 0.2 -- 2.4	& $2.26^{+0.11}_{-0.38}$ \\
		2015, \textit{XMM}				& This work	& 0.2 -- 2.4	& $2.91^{+0.16}_{-0.27}$ \\[0.25cm]
		2008, \textit{XMM}				& E15		& 0.243 -- 2.0	& $1.84$ \\
		2013-14, \textit{Chandra}		& E15		& 0.243 -- 2.0	& $1.97$ \\
		2015, \textit{XMM}				& This work	& 0.243 -- 2.0	& $2.35^{+0.16}_{-0.26}$ \\[0.25cm]
		\textit{ROSAT} All-Sky Survey	& H99, B16	& 0.1 -- 2.4	& $<12$ \\
		\hline
		\multicolumn{4}{c}{\textit{WASP-80}} \\[0.1cm]
		2014, \textit{XMM}				& S15		& 0.124 -- 2.48 & $1.6^{+0.1}_{-0.2}$ \\[0.25cm]
		2014, \textit{XMM}				& This work	& 0.2 -- 2.4	& $1.67^{+0.12}_{0.26}$ \\
		2015, \textit{XMM}				& This work	& 0.2 -- 2.4	& $1.78^{+0.11}_{0.16}$ \\
		\hline
		\end{tabular}
	\begin{tablenotes}
\item \textit{a} $10^{-14}$\flux\ (at Earth, unabsorbed)
\item References are:
		SF11: \citet{SanzForcada2011}; E15: \citet{Ehrenreich2015}; H99: \citet{Hunsch1999}; S15: \citet{Salz2015}; B16: \citet{Boller2016}.
    \end{tablenotes}
  \end{threeparttable}
\end{table}

\subsubsection{GJ\,436}
\label{sssec:GJ436}
We compare our measured fluxes to previous studies. A summary of these comparisons can be found in Table~\ref{tab:compFluxes}.

GJ\,436 previously had X-ray fluxes measured by~\citet{SanzForcada2011} and~\citet{Ehrenreich2015} (hereafter E15) using the \textit{XMM-Newton} dataset from 2008 (Obs ID: 0556560101; PI: Wheatley). The two analyses produced very different results, with the former finding the flux at Earth to be $7.3\times10^{-15}$\,\flux\ for the 0.124 -- 2.48\,keV band, almost five times smaller than the $4.6\times10^{-14}$\,\flux\ found by the latter analysis in the same energy range. We note that~\citet{Louden2017} found a similar discrepancy between their analysis and that of~\citet{SanzForcada2011} for an observation of HD\,209458. We reanalysed the previous \textit{XMM-Newton} dataset for GJ\,436 for a more direct comparison of the fluxes, obtaining a flux of $\left(2.26^{+0.11}_{-0.38}\right)\times10^{-14}$\flux\ in the 0.2 -- 2.4\,keV band. We therefore conclude that there was a modestly increased X-ray output at the time of the 2015 observations. GJ\,436 was one of the stars whose light curve was seen to vary at the 3-$\sigma$ level in section~\ref{ssec:XLC}. The difference in flux between the 2008 and 2015 datasets points to significant variation also on longer timescales.

E15 also found their analysis of the 2008 \textit{XMM-Newton} observations to agree with their Chandra data in the overlapping 0.243 -- 2.0\,keV energy range: $1.84\times10^{-14}$\flux\ versus the $1.97\times10^{-14}$\flux\ obtained when averaging across the four Chandra datasets. We measure $\left(2.35^{+0.16}_{-0.26}\right)\times10^{-14}$\flux\ in this slightly more restrictive band, again showing a modest increase on the 2008 \textit{XMM-Newton} data, but also compared to the averaged 2013-14 \textit{Chandra} data. Furthermore, we compared the emission measures of the 2015 data to the other \textit{XMM-Newton} and \textit{Chandra} observations using the method of E15 (The results for the other five datasets are plotted in their extended data fig. 8). For the most direct comparison, we fixed the temperatures and abundances to that found in E15 (i.e. \textit{not} those in Table~\ref{tab:Results}). With this method, we obtain emission measures of $9.7^{+1.3}_{-1.2}$ and $2.10^{+0.23}_{-0.22}$\,cm$^{-3}$ for the low and high temperature components, respectively. These results concur with the conclusion of E15 that there is more variation in the higher temperature component than in the soft.

We note that GJ\,436 was also observed in X-rays during the \textit{ROSAT} All-Sky Survey. \citet{Hunsch1999} reported an X-ray flux in the 0.1 -- 2.4\,keV band of $1.2\times10^{-13}$\,\flux, which is much higher than all of the other datasets. However, the revised PSPC catalog by~\citet{Boller2016} suggests the GJ\,436 detection is not real and should be treated as an upper limit. 

\subsubsection{HAT-P-11}
\citet{Morris2017} used Ca\,\textsc{ii} H \& K observations to show HAT-P-11 has an unexpectedly active chromosphere for a star of its type. Our work suggests this extends to the corona too, with its measured $L_\text{X}/L_\text{bol}$ an order of magnitude larger than that expected from W11 (Fig.~\ref{fig:measVexpt}).~\citet{Morris2017} also presented evidence for an activity cycle for HAT-P-11 in excess of 10 years using observations of chromospheric emission, with the star's S-index spending a greater proportion of its activity cycle close to maximum compared to the Sun. Despite this, our \textit{XMM-Newton} observations were taken about halfway between activity maximum and minimum, and $L_\text{X}/L_\text{bol}$ was much larger than the W11 prediction even though the star was not close to its maximum activity level.

\subsubsection{WASP-80}
WASP-80 has had a previous \textit{XMM-Newton} dataset from 2014 (Obs ID: 0744940101; PI: Salz) analysed by~\citet{Salz2015}. They reported a flux at Earth of $(1.6^{+0.1}_{-0.2})\times10^{-14}$\,\flux, in the 0.124 -- 2.48\,keV band. As for GJ\,436, we repeated the analysis of this older dataset using the same procedure as for the new observations for a more direct comparison. The fluxes can be compared in Table~\ref{tab:compFluxes}. We find a flux at Earth in the slightly more restrictive 0.2 -- 2.4\,keV band of $(1.67^{+0.12}_{-0.26})\times10^{-14}$\,\flux. This result is consistent with our observations at the newer epoch within the uncertainties.

\subsection{EUV estimation}
\label{ssec:EUVest}
In section~\ref{ssec:EUVrec}, we derived new empirical relations for reconstructing the EUV emission of stars from their observed X-rays, with the results presented in Table~\ref{tab:Results}. We now draw comparisons to past applications of other methods.

For GJ\,436, E15 obtained estimates of the EUV at 1\,au from both the C15 X-ray and~\citet{Linsky2014} Ly\,$\alpha$ methods, and found them to be remarkably similar. Adjusting for the new distance estimate from \textit{Gaia}, these were 0.92 and 0.98\,\flux, respectively. In order to procure a directly comparable flux from our own measurements, we used equation~\ref{eq:relation} (boundary energy choice \#7) from Table~\ref{tab:RelVar}. This was applied to our flux measurement from the same 2008 dataset analysed by E15 (section~\ref{sssec:GJ436}). We determine an EUV flux at 1\,au of $0.86^{+0.06}_{-0.17}$\,\flux, in satisfactory agreement with the values found by E15. The corresponding EUV flux value for the new 2015 dataset is $0.98^{+0.08}_{-0.12}$\,\flux.

\citet{Bourrier2016} also estimated the EUV flux using the~\citet{Linsky2014} method. They determine EUV fluxes of 0.88 and 0.86\,\flux\ at their two, independent epochs, in good agreement with our results from X-rays.

The MUSCLES Treasury Survey has combined observations from multiple passbands from X-ray to mid-IR to study the intrinsic spectral properties of nearby low-mass planet-hosting stars~\citep{France2016}. \citet{Youngblood2016} reconstructed the EUV flux of GJ\,436 in the 0.0136 -- 0.1\,keV band with the~\citet{Linsky2014} Ly\,$\alpha$ method, obtaining $0.83$\,\flux\ at 1\,au. Their results are therefore also consistent with extrapolation from the X-ray band.

The data presented here for HD\,97658 (Table~\ref{tab:Results}) were previously investigated by~\citet{Bourrier2017}. Unlike here, they first extrapolated to the ROSAT band, and then used C15 to extrapolate to the EUV. 
They also estimate the EUV from multiple epochs of \textit{HST} Ly\,$\alpha$ observations, applying the relations of~\citet{Linsky2014}. The results from the two methods were compatible. Our direct extrapolation to the EUV from the observed X-rays obtains an XUV flux at the planet that is marginally smaller, but consistent within the uncertainties to their best estimate. 
The agreement with EUV estimates from Ly\,$\alpha$ supports the accuracy of the two methods of reconstructing the EUV emission.

\subsection{Mass loss rates}
\label{ssec:MassLoss}

\begin{table}
	\centering
	\caption{Current mass loss rate and total lifetime mass loss estimates of the six planets is our sample for different assumed sets of $\eta$ and $\beta$. The first listed $\eta$ and $\beta$ for each planet are taken from~\citet{Salz2016}; the second is a canonical value of $\eta$ = 0.15 and $\beta$ = 1; the third provides a lower limit on the mass loss rates of these planets, motivated by Ly\,$\alpha$ observations.}
	\label{tab:ML}
	\begin{threeparttable}
	\begin{tabular}{lrrrrr} 
		\hline
		System & $\eta$ & $\beta$ & $\log \dot{M}$ & \multicolumn{2}{c}{Lifetime Loss \%} \\
		 &  &  & (g\,s$^{-1}$) & Const.$^\ast$ & J12$^\dagger$ \\
		\hline
		\multirow{3}{*}{GJ\,436} & 0.275 & 1.48 & 9.8 & 0.8 & 4.3 \\
								 & 0.15  & 1    & 9.2 & 0.2 & 1.0 \\
								 & >0.01  & 1    & >8.0 & >0.01 & >0.07  \\[0.15cm]
		\multirow{3}{*}{GJ\,3470} & 0.135 & 1.77 & 10.7 & 4.3 & 9.3 \\
								 & 0.15  & 1    & 10.2 & 1.5 & 3.5 \\
								 & >0.01  & 1    & >9.0 & >0.1 & >0.2 \\[0.15cm]
		\multirow{3}{*}{HAT-P-11} & 0.229 & 1.61 & 10.3 & 2.3 & 8.8 \\
								  & 0.15  & 1    & 9.7  & 0.6 & 2.4 \\
								 & >0.01  & 1    & >8.6 & >0.04 & >0.2 \\[0.15cm]
		\multirow{3}{*}{HD\,97658} & 0.288 & 1.75 & 9.4 & 1.7 & 3.9 \\
								   & 0.15  & 1    & 8.6 & 0.3 & 0.7 \\
								 & >0.01  & 1    & >7.4 & >0.02 & >0.05 \\[0.15cm]
		\multirow{3}{*}{HD\,149026} & 0.093 & 1.26 & 9.4 & 0.014 & 0.2 \\
									& 0.15  & 1    & 9.4 & 0.015 & 0.2 \\
								 & >0.01  & 1    & >8.2 & >0.001 & >0.01 \\[0.15cm]
		\multirow{3}{*}{WASP-80} & 0.100 & 1.24 & 10.3 & 0.004 & 0.06 \\
								 & 0.15  & 1    & 10.3 & 0.004 & 0.05 \\
								 & >0.01  & 1    & >9.2 & >0.0004 & >0.004 \\
		\hline
	\end{tabular}
    \begin{tablenotes}
\item $^\ast$ Constant lifetime XUV irradiation rate, at the current level.
\item $^\dagger$ Lifetime XUV irradiation estimated by the relations of~\citet{Jackson2012}.
    \end{tablenotes}
  \end{threeparttable}
\end{table}

We present estimated mass loss rates for all six planets in Table~\ref{tab:ML}. We follow the energy-limited approach of previous studies~\citep[e.g.][]{LDE2007,SanzForcada2011,Salz2015,Louden2017,Wheatley2017} to calculate mass loss rate estimates for each of the six systems:
\begin{equation}
\dot{M} = \frac{ \beta^2 \eta \pi F_{\text{XUV}} R^3_{\text{p}} }{ G K M_{\text{p}} },
\label{eq:massloss}
\end{equation}
where $\eta$ is the efficiency of the mass loss, $F_{\text{XUV}}$ is the total X-ray and EUV flux incident on the planet, and $\beta$ accounts for the increased size of the planetary disc absorbing XUV photons compared to visible wavelengths, equal to $R_{\text{XUV}}/R_\text{p}$. We follow the approach of~\citet{Salz2016}, outlined in their footnote 1, in using a $\beta^2$ factor~\citep{Watson1981, Lammer2003, Erkaev2007} instead of a $\beta^3$ factor~\citep[e.g.][]{Baraffe2004, SanzForcada2010}. The factor $K$, the potential energy difference between the surface and the Roche-lobe height, $R_\text{RL}$, to which material must be lifted to escape, is given by~\citep{Erkaev2007}
\begin{equation}
K = 1 - \frac{3}{2\xi} + \frac{1}{2\xi^3},
\label{eq:K}
\end{equation}
where $\xi = R_\text{RL}/R_\text{p}$. In turn, this can be approximated by $(\delta/3)^{1/3}\lambda$ where $\delta = M_\text{p}/M_*$, and $\lambda =  a/R_\text{p}$.

The value of $\eta$ for a given system has been the subject of much discussion~\citep[e.g.][and references therein]{Shematovich2014,Louden2017}, with estimates and adopted values often varying considerably from study to study~\citep[e.g.][]{Penz2008, MurrayClay2009, Owen2012}. In Table~\ref{tab:ML} we estimate mass loss rates corresponding to our observed XUV fluxes with three different assumptions for this efficiency. First, we make use of the results of coupled photoionisation-hydrodynamic simulations by~\citet{Salz2016}, which included $\eta$ and $\beta$ values for all six planets in our sample. These calculations imply relatively high mass loss efficiencies, especially for lower mass planets (Table~\ref{tab:ML}). We also include a more canonical choice of 0.15 and 1 for $\eta$ and $\beta$, respectively. These were the values adopted by~\citet{Salz2015}, allowing direct comparison of our predicted mass loss rates with those systems. Our third assumption of 1 per cent efficiency is adopted as a lower limit to the likely mass loss efficiency, and hence mass loss rates, motivated by observational constraints from contemporaneous measurements of the XUV irradiation and resulting mass loss detected through Ly\,$\alpha$ absorption in individual systems \citep[e.g.][]{Ehrenreich2011}. For GJ\,436b an efficiency as low as 0.5 per cent has been shown to be sufficient to explain the observed strong Ly\,$\alpha$ absorption, if the material is completely neutral as it leaves the planet~\citep{Ehrenreich2015,Bourrier2016}. For the hot Jupiter HD\,189733b a similarly low lower limit of 1 per cent is also sufficient to explain the observed absorption by H\,\textsc{i}, although a somewhat higher efficiency is likely to be needed to account for the unobserved ionised hydgrogen~\citep{LDE2012}. For the super-Earth HD\,97658b, upper limits on Ly\,$\alpha$ absorption from~\cite{Bourrier2017} suggest a mass loss efficiency that could be substantially lower than that predicted by~\citet{Salz2016}, depending on the ionisation fraction of material leaving the planet. Since this fraction is poorly known, the assumed value of 1 per cent efficiency in Table~\ref{tab:ML} provides a lower limit on the mass loss rates of the planets. The true efficiency is likely to be higher, and indeed a much higher mass loss efficiency is also required for HD\,209458~\citep{Louden2017}. Given this uncertainty in the mass loss efficiencies, we present mass loss rates for all three choices of $\eta$ and $\beta$ in Table~\ref{tab:ML}.

Following~\citet{Salz2016}, the mass loss rate estimates for GJ\,436b and HD\,97658b exceed the values derived by modelling Ly\,$\alpha$ observations with the EVaporating Exoplanets (EVE) code~\citep{Bourrier2016,Bourrier2017}. The resulting mass loss estimates for the other choices of $\eta$ and $\beta$ for these planets are both lower and closer to their respective estimates from Ly\,$\alpha$, although the $\eta = 0.01$ results perhaps provide a slight underestimation.

As discussed by~\citet{Owen2016}, EUV-driven evaporation of close-in planets can be in one of three regimes: energy-limited, recombination-limited, and photon-limited. Their numerical calculations show that the transition between the three regimes does not occur at a single point, rather over a few orders of magnitude. However, their fig. 1 allows us to determine that GJ\,3470b, HAT-P-11b, and HD\,97658b are likely in the region of energy-limited escape. HD\,149026b and WASP-80b lie close to the transition between the energy-limited and recombination-limited regions. Note that energy conservation always applies in the planetary thermospheres, but in the case of recombination-limited escape, a larger fraction of the absorbed radiative energy is re-emitted by recombination processes, so that less energy is available to drive the planetary wind. Therefore, the recombination-limited regime exhibits lower evaporation efficiencies than the energy-limited regime. In agreement with their intermediate location close to the recombination regime, the estimates of $\eta$ for HD\,149026b and WASP-80b from \citet{Salz2016} are smaller than for the other four planets.

\subsubsection{Total lifetime mass loss}
\citet{Jackson2012} produced a set of relations characterising the evolution of the X-ray emission with stellar age. As a result, they were able to further derive relations that can be used to estimate the total X-ray emission of a star over its lifetime to date. In turn, this could be used to estimate the total mass lost from an exoplanet. 
This would be particularly useful to apply to close-in super-Earth and mini-Neptune-sized planets, to investigate if they could have suffered substantial or total loss of a gaseous envelope. For middle-aged systems, if this happened, it is likely to have occurred much earlier in their life when the coronal emission of their host was much greater.

We apply equation 8 of~\citet{Jackson2012}, together with the ages from Table~\ref{tab:SysParam}, in order to estimate the lifetime X-ray output from each of the six host stars in our sample. The results are given in Table~\ref{tab:ML}. Additionally, we consider the corresponding EUV by applying relation \#1 (Table~\ref{tab:RelVar}) to the estimated X-ray output at 1000\,yr steps and integrating over the resulting lifetime evolution. We then scale the results to the average orbital separation of the system's planet, and apply equation~\ref{eq:massloss} to estimate the total mass lost over the planet's lifetime. Estimates for all three sets of choices of $\eta$ and $\beta$ are included. Also in Table~\ref{tab:ML} are estimates for the total percentage mass loss over the lifetime of each planet, assuming a constant XUV irradiation rate, at the current level. While we assume a constant radius across the planet's lifetime, if substantial evolution has occurred, the use of a constant radius could mask a greater total lifetime mass loss than our estimates~\citep{Howe2015}.

The lifetime loss results are sensitive to the assumed $\eta$ and $\beta$, as well as discrepancies between the theoretically expected $L_\text{X}/L_\text{bol}$ and that observed. Additionally, HD\,149026's subgiant nature will affect its estimate. However, more qualitatively, the four smallest planets studied are expected to have lost a much greater percentage of their mass over their lifetime than the other two much larger planets in the sample.

Applying equation~\ref{eq:massloss} to a planet of Neptune mass and radius with the same irradiation history as HD\,97658b, we find such a planet would have lost $\sim$3.5 per cent of its mass over its lifetime. This is in contrast to closer-in planets like CoRoT-7b, which is suspected to have suffered a near-complete loss of its gaseous envelope due to intense irradiation~\citep{Jackson2010}.

\subsection{Ly\,$\alpha$ estimation}
Ly\,$\alpha$ observation of highly irradiated exoplanets is an important tool to determine the extent of atmospheric evaporation. Ly\,$\alpha$ transits have proven successful in detecting evaporating atmospheres. Additionally, as previously stated, Ly\,$\alpha$ observations also provide a separate regime from which EUV reconstruction can be performed.

For each of the systems in our sample, we have estimated the Ly\,$\alpha$ output in two steps. Firstly, we used equation~\ref{eq:relation} (boundary relation \#7) to calculate the EUV flux in the 0.0136 -- 0.124\,keV band. Then, we applied the relations of~\citet{Linsky2014}, linking Ly\,$\alpha$ and EUV fluxes at 1\,au. By plotting the curves given by the relations, we approximated the Ly\,$\alpha$ flux according to the position of each systems' EUV estimation. Table~\ref{tab:Results} gives Ly\,$\alpha$ luminosity, $L_{\text{Ly}\alpha}$, estimates for our six systems, and the corresponding flux at Earth, $F_{\text{Ly}\alpha, \oplus}$. For GJ\,436 and HD\,97658, we additionally include literature values. While the results from~\citet{Bourrier2016} and~\citet{Bourrier2017} for GJ 436 are remarkably consistent with our results, there is less agreement with those of~\citep{Youngblood2016} for HD 97658, although
their value is poorly constrained with larger errors.

Our analysis suggests that the HAT-P-11 system is the best candidate for Ly\,$\alpha$ observations, of those that have not previously been studied in this way. We predict the star to have the largest apparent Ly\,$\alpha$ brightness of the three, while we estimate the planet's mass loss rate to be larger than that of GJ\,436b by about a factor of three. This is largely because the observed X-ray flux is significantly higher than expected. While our $F_{\text{Ly}\alpha, \oplus}$ prediction does account for interstellar absorption, the Ly\,$\alpha$ snapshot of WASP-80 by~\citet{SalzThesis} shows that large transits could even be detected for one of the most distant systems in this sample. Hence, all of the studied systems likely qualify for systematic Ly\,$\alpha$ transit observations, but HAT-P-11 and GJ\,3470 appear to be the best suited.

\subsection{WASP-80 NUV transit}
The OM light curve of WASP-80 allowed us to detect the planetary transit in the near ultraviolet. Our best fit $R_p / R_*$ of $0.125^{+0.029}_{-0.039}$ corresponds to a NUV transit depth of $1.6^{+0.5}_{-0.7}$ per cent, and a planet radius of $0.69^{+0.16}_{-0.22}$\,R$_{\text{J}}$. In comparison, the discovery paper reported a visible light $R_p / R_*$ of $0.17126^{+0.00031}_{-0.00026}$~\citep{Triaud2013}, while~\citet{Mancini2014} measured $0.17058\pm0.00057$, and~\citet{Kirk2018} found $0.17113\pm0.00138$. The latter study also found little evidence of large variation in the radius of WASP-80\,b's across the visible and near infrared. Our results are consistent, though the best fit transit is shallower by 1.59-$\sigma$. This is perhaps a hint that the NUV transit is shallower. It would be desirable to follow up with more observations in the NUV that could constrain the depth to a higher precision, particularly given the size of the uncertainties on our fitted depth.

A shallower NUV transit would not be without precedent. With ground-based observations, \citet{Turner2016} found smaller NUV (U band) transit depths for hot Jupiters WASP-1b and WASP-36b with significance 3.6-$\sigma$ and 2.6-$\sigma$, respectively. Physically, a shallower transit in NUV could result from the planet passing in front of dimmer regions of the star. The contrast between the areas of the stellar disc the planet crosses and brighter regions elsewhere would also need to be higher in the NUV than visible light for this explanation to be feasible. Unocculted faculae could possibly produce this effect. Spectral modelling of faculae have shown the contrast in intensity between the facula and elsewhere on the stellar disc is greater in the UV than in the visible and IR, as well as for regions closer to the limb of the disc~\citep[e.g.][]{Unruh1999,Norris2017}. Indeed, stellar activity in the transit light curve of WASP-52\,b was interpreted by~\citet{Kirk2016} as occulted faculae. WASP-80\,b has a much lower impact parameter than WASP-52\,b, and so spends less time crossing regions close to the limb, making it more likely that high-contrast faculae close to the limb would go unocculted.

\section{Conclusions}
\label{sec:conclusions}
We have analysed \textit{XMM-Newton} data to investigate the XUV environments of six nearby transiting planets that orbit in close proximity to their host star, ranging in size from Jupiter-size to super-Earth. For each star, we directly measure the flux in the 0.2 -- 2.4\,keV band by fitting a two temperature APEC model. We use a similar approach to~\citet{Chadney2015} in using Solar \textit{TIMED/SEE} data to derive a new set of relations for reconstructing the unobservable EUV emission. We use different boundary choices between the EUV and X-ray bands based on the current generation of X-ray instruments. The resulting estimates for the full XUV range of GJ\,436 and HD\,97658 are in good agreement with past reconstructions from X-ray and Ly\,$\alpha$.

With the contemporaneous measurements from the OM in the near ultraviolet, we searched for transits in the fast mode data. We successfully uncovered a transit from OM data for the first time. Our resulting fit showed a best fit transit depth for WASP-80b consistent with previous studies in visible light and in the near infrared within the uncertainties. However, there is a hint that the depth could be shallower, and so we recommend further observations in the NUV to investigate more precisely the possibility of a smaller transit depth at these wavelengths.

We investigated how our measured X-ray emission, and its ratio to the corresponding bolometric luminosity, compared to that expected from the known rotation rate and estimated Rossby number of each star. We see a possible trend to slower rotating stars being brighter than expected. The scatter in these results highlights the importance of investigating systems of interest with dedicated observations.

The mass loss rate for each planet was estimated. Our mass loss rates for GJ\,436b and HD\,97658b calculated using the efficiency and absorption radii determined by~\citet{Salz2016} appear inconsistent with analysis of Ly\,$\alpha$ observations. Based on our Ly\,$\alpha$ emission estimates, all six systems qualify for observations at those wavelengths. However, HAT-P-11b and GJ\,3470b are best suited of the four without previous extensive investigation due to their proximity to the Solar System. Both systems have larger predicted mass loss rates than GJ\,436b or HD\,97658b. Finally, we determine that the super-Earth and three Neptunes among our sample are likely to have lost a larger mass fraction over their lifetimes than the other two larger planets.

\section*{Acknowledgements}

We thank the referee, Jeffrey Linsky, for his helpful comments which improved the quality of this manuscript. We thank Don Woodraska for providing some clarifications with the TIMED/SEE data, and Jorge Sanz-Forcada for providing the synthetic spectra data for AD\,Leo and AU\,Mic. G.W.K. and J.K. are supported in part by STFC studentships. P.J.W. is supported by an STFC consolidated grant (ST/P000495/1). MS acknowledges support by the DFG SCHM 1032/57-1 and DLR 50OR1710. D.E. and V.B. acknowledge support from the National Centre for Competence in Research `PlanetS' of the Swiss National Science Foundation (SNSF) and from the European Research Council (ERC) under the European Union's Horizon 2020 research and innovation programme (project `{\sc Four Aces}'; grant agreement No 724427).




\bibliographystyle{mnras}
\bibliography{paper_clean} 




\appendix

\section{High precision fast photometry with the \textit{XMM-Newton} Optical Monitor}
\label{Appen:OM}

\begin{figure}
 \includegraphics[width=\columnwidth]{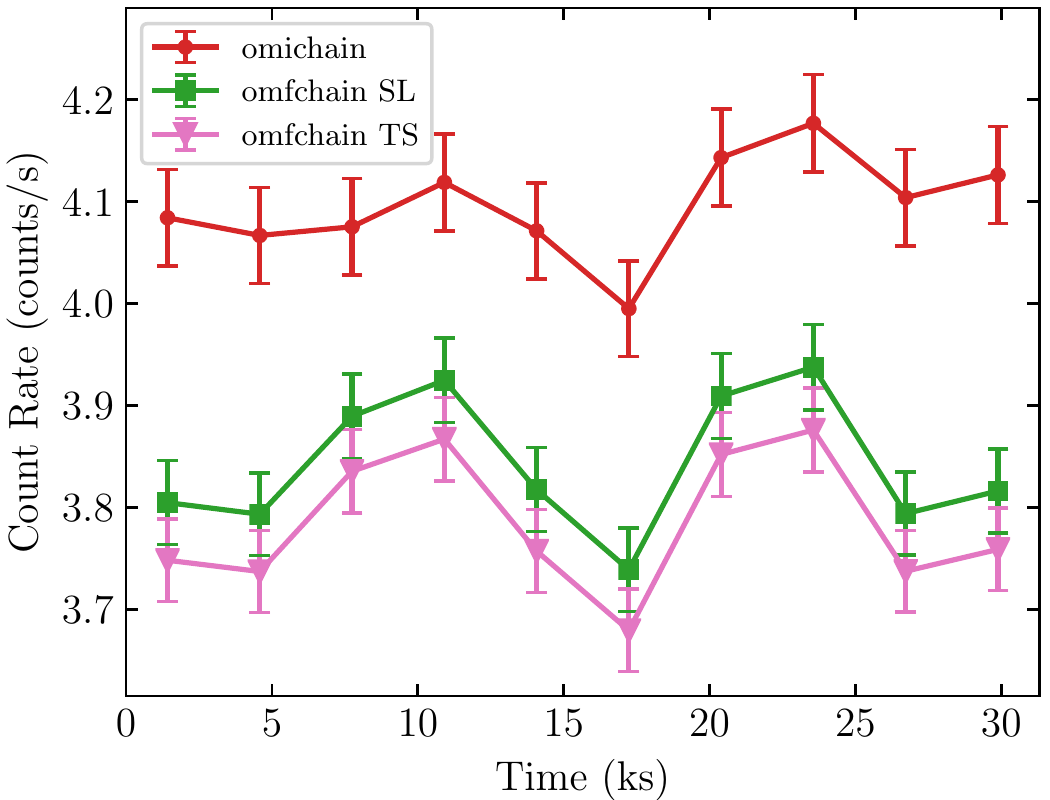}
 \caption{Comparison of the \textit{XMM-Newton} Optical Monitor light curves for WASP-80. The image mode data reduced by \textsc{omichain} is shown by the red circles. Two fast mode light curves from the \textsc{omfchain} outputs are displayed: one taken from the source list (SL) for each overall exposure (green squares), the other from the time series (TS) binned to the same cadence (pink triangles).}
 \label{fig:chainComp}
\end{figure}

In assessing the OM data for WASP-80, we noticed the outputs from the standard \textsc{sas} analysis chains for the image and fast mode data, \textsc{omichain} and \textsc{omfchain}, respectively, did not fully agree with each other. Fig.~\ref{fig:chainComp} highlights the differences between the shape of the image mode light curve (red circles) and fast mode light curves (green squares and pink triangles; where the green squares are from the per-exposure source lists accompanying the fast mode light curve, and the pink triangles represent the fast mode time series binned to the same cadence, both from \textsc{omfchain}). The most obvious differences are the jump after the first two points and drop down before the last two points.

Our hypothesis for the cause of the discrepancy between the image and fast mode chains was that this was due to the different source apertures employed. \textsc{omichain} uses 12 pixel radii apertures for the image mode data, but \textsc{omfchain} uses only 6 pixel radii regions because of the small size of the fast mode window. Unfortunately, we could not test this hypothesis using the analysis chains. The aperture size used by \textsc{omichain} is not able to be modified, and although the sizes employed by \textsc{omfchain} are customisable, the fast mode window is far too small for apertures with a radius of 12 pixels to be used. Therefore, to test our hypothesis, we instead analysed the data using a standard photometry code.

\begin{figure}
 \includegraphics[width=\columnwidth]{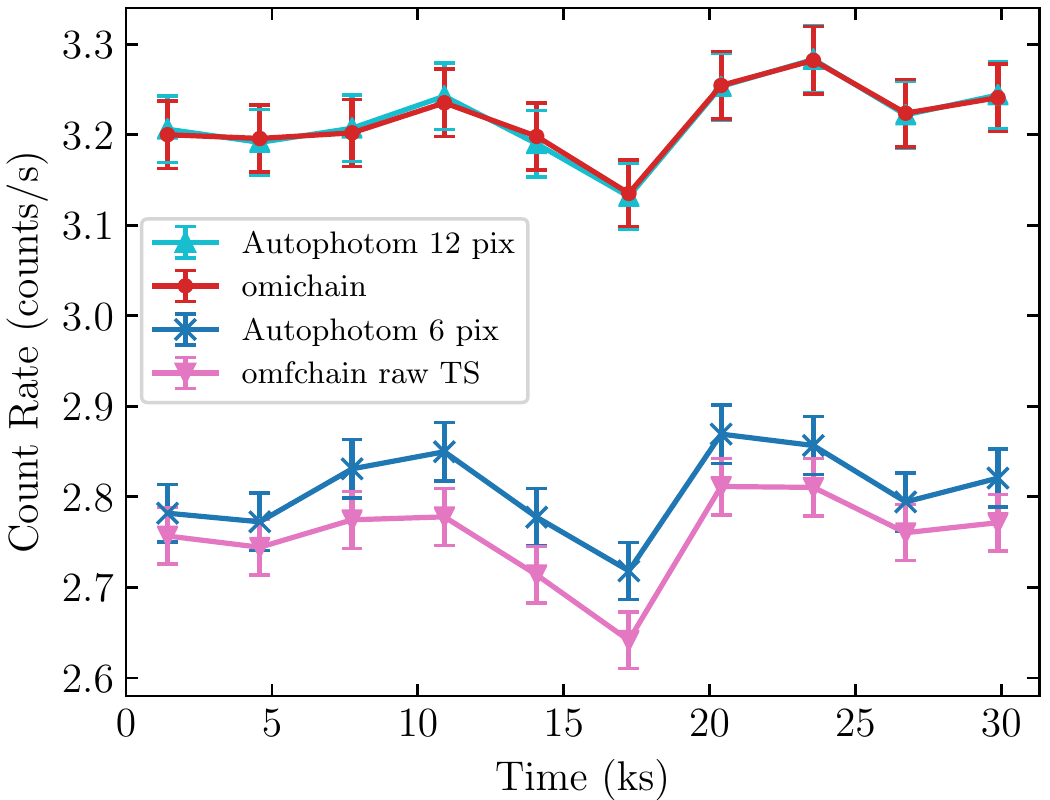}
 \caption{Comparison of the raw \textsc{omichain} (red circles) and \textsc{omfchain} time series (pink down-pointing triangles) with \textsc{autophotom} analyses using 12 (cyan up-pointing triangles) and 6 (blue crosses) pixel radii apertures. This shows the main difference between \textsc{omichain} and \textsc{omfchain} light curves is due to the different extraction radii used.}
 \label{fig:chainAutoTS}
\end{figure}

We performed aperture photometry on the image mode data using the \textsc{autophotom} routine, part of the \textsc{photom} package~\citep{Eaton2009} from the \textsc{starlink} project~\citep{Currie2014}. This was done using source aperture radii of 12 and 6 pixels. These light curves, along with the raw light curves from the \textsc{omichain} and \textsc{omfchain}, are displayed in Fig.~\ref{fig:chainAutoTS}. Our 12 pixel aperture extraction using \textsc{autophotom} (shown as cyan up-pointing triangles) is in excellent agreement with the \textsc{omichain} light curve (red circles), and our 6 pixel aperture \textsc{autophotom} extraction is very similar in shape to the raw \textsc{omfchain} time series. This confirms our hypothesis that the main difference between \textsc{omichain} and \textsc{omfchain} can be attributed to the different extraction radii. However, there is slight difference in shape towards the middle of the observation, which points to a second effect (there is also an offset similar to that seen between the two \textsc{omfchain} outputs in Fig.~\ref{fig:chainComp}). 

\begin{figure}
 \includegraphics[width=0.76\columnwidth]{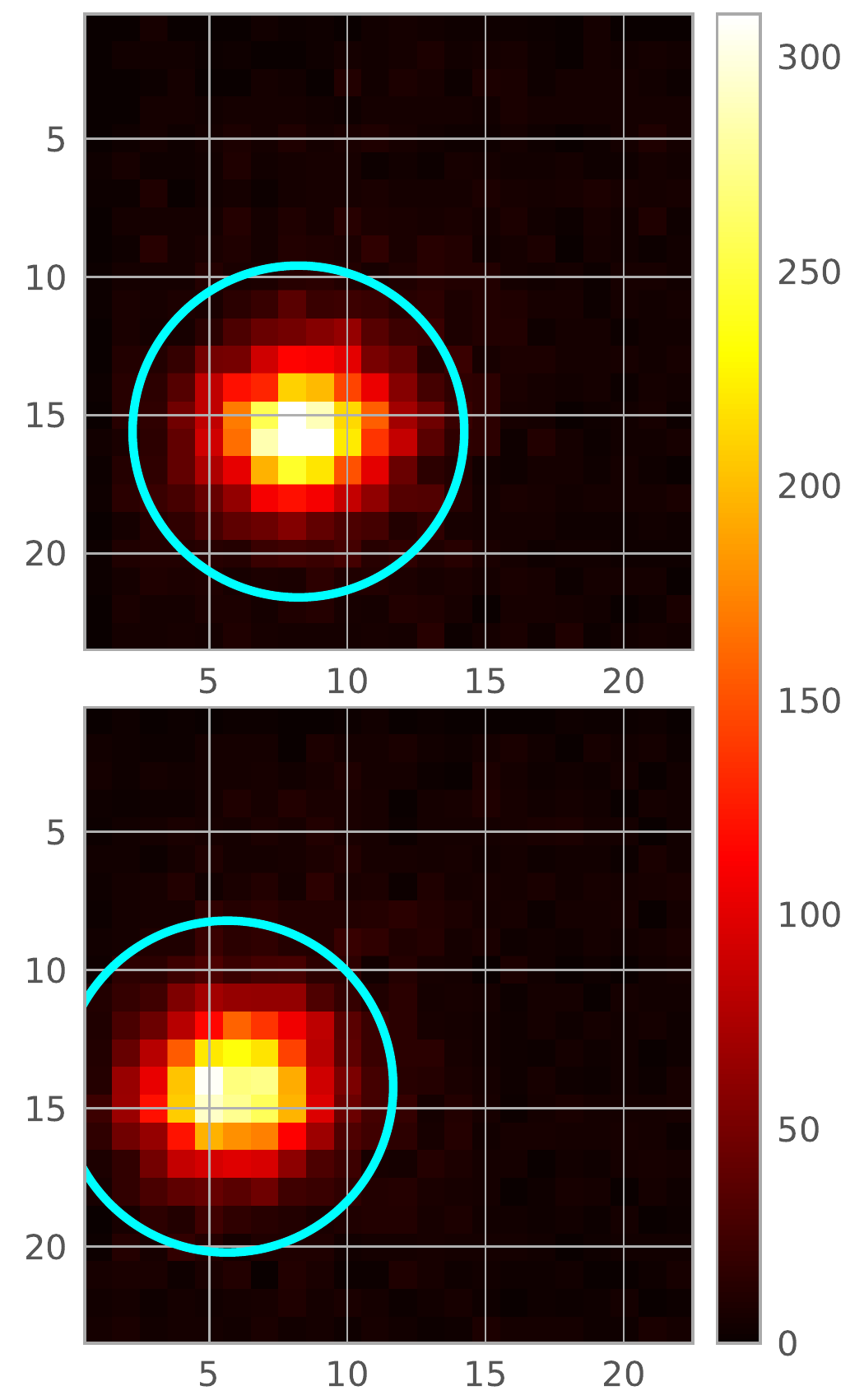}
 \caption{Comparison of two fast mode exposures, and how the source position within the window changes. For exposure 401 (top panel), the \text{omfchain} aperture used to extract the time series, overplotted in cyan, remains fully within the window. However, the aperture runs into the side of the window in exposure 007 (bottom panel), causing a small discrepancy with the corresponding image mode data when a same-sized aperture is used.}
 \label{fig:fastmodewindow}
\end{figure}

We believe this second effect is the result of the source moving in the fast mode window, causing the extraction aperture to extend a little beyond the fast mode window for these exposures. This is highlighted in Fig.~\ref{fig:fastmodewindow}, which shows two fast mode window exposures: `401' and `007'. The former is unaffected by this issue, whereas the latter is the worst afflicted. The points with a greater offset in the fast mode comparison in Fig.~\ref{fig:chainAutoTS} correspond to the exposures where the PSF runs into the edge of the fast mode window.

We conclude that the differences in shape we see in Fig.~\ref{fig:chainComp} can be understood as primarily resulting from the different aperture sizes used, with a further, smaller contribution from the source aperture running into the sides of the fast mode window. Therefore, we feel justified in correcting fast mode data from \textsc{omfchain} by the corresponding image mode data from \textsc{omichain}. Taking the ratio of the image mode data to the fast mode time series binned to the same cadence (i.e. the ratio of the red and pink light curves in Fig.~\ref{fig:chainComp}) provides a suitable correction. Each individual time bin in our analysis in Section~\ref{ssec:fastmodeOM} was therefore multiplied by this ratio, as calculated for the corresponding exposure.

\bsp	
\label{lastpage}
\end{document}